\documentclass[a4paper,12pt]{article}

\usepackage[T1]{fontenc} 

\usepackage{draft}
\usepackage{putex} 

\usepackage{cite}

\usepackage[
      colorlinks=true,
      linkcolor=black,
      urlcolor=blue,
      filecolor=black,
      citecolor=red,
      ]{hyperref}
      
\usepackage{graphicx}
\usepackage{xcolor}
\usepackage{amsfonts,amsmath,amssymb,bm}
\usepackage{verbatim}
\usepackage{array}
\usepackage{mathrsfs}
\usepackage{mathtools}
\usepackage{yfonts}
\usepackage{dsfont}
\usepackage{bbm}
\usepackage{colonequals}
\usepackage{amscd}
\usepackage{slashed}

\usepackage{float}
\usepackage{afterpage}

\usepackage{subcaption}

\usepackage{wrapfig}

\usepackage{suffix,mathtools,cancel,bbm}

\usepackage{multirow}

\makeatletter
\def\leftrightarrowsfill@{\arrowfill@\leftrarrows\Rrelbar\lrightarrows}
\newcommand{\xleftrightarrows}[2][]{\ext@arrow 3399\leftrightarrowsfill@{#1}{#2}}
\makeatother

\begin{document}

\title{Trace relations and open string vacua}

\authors{Ji Hoon Lee}

\institution{PI}{Perimeter Institute for Theoretical Physics, Waterloo, Ontario, Canada N2L 2Y5}

\abstract{We study to what extent, and in what form, the notion of gauge-string duality may persist at finite $N$. It is shown, in the half-BPS sector, that the states of D3 giant graviton branes in $\mathrm{AdS}_5 \times S^5$ are holographically dual to certain auxiliary ghosts that compensate for finite $N$ trace relations in $U(N)$ $\mathcal{N}=4$ super Yang-Mills. The complex formed from spaces of states of bulk D3 giants is observed to furnish a BRST-like resolution of the half-BPS Hilbert space of $U(N)$ $\mathcal{N}=4$ SYM at finite $N$. We argue that the identification between the states of certain bulk D-branes and the auxiliary ghosts in the boundary holds rather generally at vanishing 't Hooft coupling $\lambda = 0$. We propose that a complex, which furnishes a BRST-like resolution of the finite $N$ Hilbert space of a boundary $U(N)$ gauge theory at $\lambda = 0$, should be identified as the space of states of the dual string theory in the $\alpha' \to \infty$ limit. The Lefschetz trace formula provides the holographic map in this regime, where bulk observables are computed by taking the alternating sum of the expectation values in an ensemble of states built on each open string vacuum. The giant graviton expansion is recovered and generalized in a limit of the resolution.}
\date{}

\maketitle

\tableofcontents

\pagebreak

\setcounter{page}{1}

\section{Introduction} \label{sec:intro}

A central theme in our study of holography and gauge-string duality is the idea that the Feynman diagrams of a large $N$ gauge theory reorganize into the genus expansion of \textit{some} string theory \cite{tHooft:1973alw,Maldacena:1997re,Witten:1998qj,Gubser:1998bc}. A particularly striking aspect of 't Hooft's proposal is its generality: any theory of gauged large $N$ matrices exhibits such a topological expansion at weak coupling.

This work investigates to what extent, and in what form, the notion of gauge-string duality may persist at finite $N$. 

At finite $N$, one is confronted with conceptual issues regarding the spectrum of the bulk and boundary theories related via holographic duality. In the boundary gauge theory, the total number of states---at fixed values of the conserved charges and the 't Hooft coupling $\lambda$---decreases as the rank $N$ of the gauge group is decreased. This occurs because states formed by exciting the vacuum with single trace operators are orthogonal at large $N$ but become constrained by trace relations when $N$ is finite. On the other hand, in the bulk string theory, non-perturbative states such as black holes and D-branes become lighter at finite string coupling $g_s$. Such objects, which are present in addition to the perturbative string states, would naively result in many more states at finite $g_s$ compared to those at small $g_s$, at fixed charges and $L^4/\alpha'^2$.

How does one reconcile these differing expectations? The correct resolution of the apparent conflict is to posit that non-perturbative objects in string theory represent highly redundant descriptions of the same set of quantum states: a state composed of $N$ fundamental strings may admit a description as a brane, and a state composed of $N$ D-branes admits a description in terms of the backreacted geometry. The intuition that non-perturbative effects in string theory place an upper bound on the single-string spectrum is known as the stringy exclusion principle \cite{Maldacena:1998bw,Myers:1999ps,McGreevy:2000cw}. While this principle is thought to be a generic feature of string theory, there is little quantitative understanding of the mechanism underlying the principle.

The giant graviton expansion, a recently-proposed formula that relates the BPS sectors of gauge theories and their string duals, suggests how the stringy exclusion principle is realized in the spectrum of the bulk string field theory \cite{Gaiotto:2021xce,Arai:2019xmp,Imamura:2021ytr}:\footnote{There exist different but compatible versions of the formula that are related by wall-crossing \cite{Gaiotto:2021xce,Lee:2022vig}. We write a simple form of the formula which is a special case of the relation proposed in the current work.}
\begin{equation} \label{eq:ggeintro}
Z_N(q) = Z_\infty(q) \sum_{k=0}^\infty q^{k N} \hat{Z}_{k} (q).
\end{equation}
The formula expresses $Z_N(q)$, the superconformal index \cite{Kinney:2005ej} of a $U(N)$ gauge theory, as a sum over BPS indices $q^{k N} Z_\infty(q) \hat{Z}_{k} (q)$ of the bulk string theory on backgrounds labelled by $k$ giant graviton branes. $Z_\infty(q)$ is the index of Kaluza-Klein modes and $q^{k N} \hat{Z}_{k} (q)$ is the index of $k$ giant graviton branes and their open-string excitations. Giant gravitons are BPS branes on $\mathbb{R} \times S^{m-2} \subset \AdS_n \times S^m$ that are supported via a large angular momentum induced from the background Ramond-Ramond potential \cite{McGreevy:2000cw,Grisaru:2000zn,Hashimoto:2000zp}. See, for example, \cite{Lee:2022vig,Arai:2020qaj,Arai:2020uwd,Murthy:2022ien,Imamura:2022aua,Choi:2022ovw,Liu:2022olj,Eniceicu:2023uvd,Gautason:2023igo,Beccaria:2023zjw,Beccaria:2023hip,Beccaria:2023sph,Fujiwara:2023bdc} for related works.

A salient feature of the giant graviton expansion is the absence of explicit contributions from gravitational saddles, even in the $\frac{1}{16}$-BPS sector of IIB string theory and $\mathcal{N}=4$ super Yang-Mills where supersymmetric black holes are present. That the asymptotic degeneracies of bulk configurations of strings and branes in \eqref{eq:ggeintro} reproduce the entropy of large BPS black holes in $\AdStimesS$ was demonstrated recently \cite{Choi:2022ovw,Beccaria:2023hip}. 

The following property of the formula realizes the stringy exclusion principle: BPS spectra on string backgrounds labelled by different numbers of giant graviton branes exhibit large cancellations to reproduce the finite $N$ index. This occurs even in sectors of gauge theory that are purely bosonic, such the half-BPS sector of $\Nfour$ SYM:
\begin{equation} \label{eq:gge half-BPS intro}
\frac{1}{\prod_{n=1}^N (1-x^n)} = \frac{1}{\prod_{n=1}^\infty (1-x^n)} \sum_{k=0}^\infty (-1)^k x^{k N} \frac{x^{k(k+1)/2}}{\prod_{m=1}^k (1 - x^m)}.
\end{equation}
Observe that each summand, supplemented with the half-BPS Kaluza-Klein spectrum, overcounts the half-BPS spectrum of $U(N)$ $\Nfour$ SYM. The overcounting is saved only by the overall minus signs $(-1)^k$ that appear with odd numbers of D3 giants. Motivated by this observation, we suggested in \cite{Lee:2022vig} that the bulk BPS Hilbert space possesses an extra $\mathbb{Z}_2$-grading, on top of the usual fermion number grading $(-1)^F$ in the gauge theory, that labels open string vacua associated to different D-brane backgrounds.

Natural questions are then, what is the bulk origin of this grading? How does the $U(N)$ gauge theory reflect the extra bulk grading if the grading is not manifest in its physical Hilbert space? Most importantly, how is the presence of the extra grading consistent with holography?

In Section \ref{sec:dbraneghost}, we show in the half-BPS sector that the space of states of $k$ giant graviton branes on $\AdStimesS$ is holographically dual to the space $\mathcal{V}_k$ of certain ``heavy'' ghosts in $\Nfour$ SYM that compensate for null states due to finite $N$ trace relations, with total ghost number $k$. That is, \textit{open strings on bulk D-branes are not dual to any physical states of a $U(N)$ gauge theory, but rather are objects that systematically cancel out the redundancies among states that arise at finite $N$}. We find that normalizable states of the worldvolume theory of $k$ D3 giants are gauge-invariant differential $k$-forms constructed from worldvolume scalars and that such states survive in the large curvature limit of $\AdStimesS$ due to a localization mechanism. We identify the form degree $k$ of these states as the extra grading in the bulk that labels open string vacua. The spectrum of normalizable states of half-BPS D3 giants are recovered by that of finite $N$ trace relations, whose generating function is the determinant operator $\det(z - X)$ in the gauge theory. Due to the grading, the normalizable states are identified with auxiliary heavy ghosts that compensate for finite $N$ null states.\footnote{There exist previous proposals for the holographic dual of open strings on D3 giants in terms of physical operators in the boundary \cite{Balasubramanian:2002sa,Balasubramanian:2004nb,deMelloKoch:2007rqf}. Note, however, that the spectrum of \textit{normalizable} excitations of half-BPS D3 giants are reproduced by the spectrum of auxiliary ghosts that compensate for trace relations.}

We use the term \textit{D-brane states} to refer to the states of bulk D-branes obtained by quantizing the open string fields.

Based on the above identification, we show that the spaces of half-BPS D-brane states form a complex that furnishes a BRST-like resolution of the half-BPS Hilbert space $\cH_N$ of $U(N)$ $\Nfour$ SYM at finite $N$. The resolution indicates that open strings on bulk D-branes are encoded in the \textit{relations} between the single-trace generators that arise in the boundary gauge theory at finite $N$. The giant graviton expansion \eqref{eq:gge half-BPS intro} in the half-BPS sector is recovered using the fact that the alternating sum of the Hilbert series of modules in a resolution vanishes. Thus, we identify the complex formed from spaces of D-brane states as the half-BPS Hilbert space of IIB string theory in the large curvature limit of $\AdStimesS$.

While the bulk computations in the half-BPS sector rely on supersymmetry and the presence of a weakly-curved gravity limit, we argue that the identification between D-brane states and heavy ghosts is quite general and holds independently of those properties. Trace relations between gauge-invariant operators are present in any finite $N$ gauge theory with adjoint fields, and such a theory may be rewritten as a $N=\infty$ gauge theory supplemented with auxiliary ghosts for trace relations that are present in the original theory, at least when $\lambda = 0$.

Let us state our main proposal here, leaving a review of the mathematical background to Section \ref{sec:bulkhilbertspace}: Consider a four-dimensional $U(N)$ gauge theory on $\mathbb{R} \times S^3$ possessing only fields in the adjoint representation of the gauge group, at zero 't Hooft coupling $\lambda = 0$. Let $R$ be the ring of polynomials of traces of fields and their derivatives (before accounting for trace relations). The space of states $\cH_N$ of the gauge theory is the quotient of $\cH_\infty = R |0\rangle$ by the submodule generated by the set of all trace relations at a given value of $N$.

\textit{Our proposal is that the complex $\cV^N$:}
\begin{equation}
\cdots \ \rightarrow \ \cV_3^{N} \ \xrightarrow{\hQ} \ \cV_2^{N} \ \xrightarrow{\hQ} \ \cV_1^{N} \ \xrightarrow{\hQ} \ \cH_\infty \rightarrow 0,
\end{equation}
\textit{which furnishes a Koszul-Tate resolution $\cK$ of $\cH_N$ over the ring $R$, should be identified as the space of states of the dual string theory in the $\alpha' \to \infty$ limit.} The homology of $\cV^N$ is concentrated at heavy ghost number $0$ and coincides with $\cH_N = \cH_\infty / \hQ \cV_1^N$. The free $R$-module $\cV_k^N$ is, in a sense that is elaborated, the $\alpha' \to \infty$ limit of the bulk space of states on a string background with $k$ wrapped D-branes. The observables of the $U(N)$ gauge theory are related to those of its string dual via the Lefschetz trace formula
\begin{equation}
\Tr_{\cH_N} \cO = \sum_{k=0}^\infty (-1)^k \Tr_{\cV_k^N} \cO
\end{equation}
which provides the holographic map in the regime of $\lambda = 0$ and finite $N$. Bulk observables are computed by taking the alternating sum of the expectation values in an ensemble of states built on each open string vacuum.

In Section \ref{subsubsec:largeN}, we show how the giant graviton expansion of \cite{Gaiotto:2021xce} is recovered and generalized in a certain limit of the complex $\cV^N$. In Section \ref{subsubsec:lowerbound}, we place a lower bound on the degeneracies of D-brane states using the uniqueness of minimal resolutions. In Section \ref{sec:basicexamples}, we discuss Koszul-Tate resolutions in simple physical examples.

Many questions are left open in our work. For instance, while our proposal suggests how gauge-string duality at weak 't Hooft coupling can be generalized to the regime of finite $N$, there remains the difficult problem of finding a consistent bulk string theory with the features suggested by our proposal. We hope that this work contributes to the understanding of some kinematical properties of this theory.

We conclude with a partial list of open questions in the Discussion.

\section{D-branes as heavy ghosts} \label{sec:dbraneghost}

In this section, we argue that the normalizable states of half-BPS D3 giant gravitons in the bulk are holographically dual to certain auxiliary ``ghosts'' in $U(N)$ $\Nfour$ super Yang-Mills that compensate for null states due to finite $N$ trace relations. That is, open strings on D3 giants are not dual to any physical states of the $U(N)$ gauge theory, but rather are objects that systematically cancel out the redundancies among states that arise at finite $N$.

Indeed, giant gravitons were conceived in order to provide a bulk explanation for the truncation of the single trace spectrum at finite $N$, in terms of the polarization of large charge gravitons into branes. The goal in this section is to understand precisely why the identification between open strings on giants and ghosts for trace relations should be made.

\subsection{A twist on the worldvolume} \label{subsec:twistworldvolume}

Let us show that the background Ramond-Ramond potential induces a twist of the stress tensor on the worldvolume of a D3 giant. The sign of the twist on the D3 worldvolume will later have an important consequence: normalizable states that describe transverse fluctuations of branes in $S^5$ become valued in gauge-invariant combinations of differential forms constructed from worldvolume scalars.

Let us work in $\AdS_5 \times S^5$ coordinates
\begin{equation}
ds^2 = -\left( 1 + \frac{\rho^2}{L^2} \right) dt^2 + \left( 1 + \frac{\rho^2}{L^2} \right)^{-1} d\rho^2 + \rho^2 d\hat{\Omega}_{S^3}^2 + \frac{L^2}{L^2 - r^2} dr^2 + r^2 d\theta^2 + (L^2 - r^2) d\Omega_{S^3}^2
\end{equation}
covering half of the five-sphere, with Ramond-Ramond 4-form potential
\begin{equation}
C_4 = \frac{\rho^4}{L} dt \wedge d\hat{\Omega}_{S^3} + (L^2 - r^2)^2 d\theta \wedge d\Omega_{S^3}.
\end{equation}
The coordinate $r \in [0, L)$ is the radius on the transverse plane to $S^3$ inside $S^5$. The maximal D3 solution sits at $r=0$ with $\dot\theta = \frac{1}{L}$ and has energy $E = \frac{N}{L}$ and angular momentum $R = N$.\footnote{The angular velocity at $r=0$ arises because the maximal D3 solution is a limit of stable solutions as $r \to 0$ and $\dot\theta \to \frac{1}{L}$.} Despite wrapping a topologically trivial cycle, giant gravitons are stable BPS solutions because they possess a large angular momentum induced from the background Ramond-Ramond potential $C_4$.

We are interested in half-BPS transverse fluctuations of a giant on a maximal $S^3 \subset S^5$ fixed under the $U(1)_X$ isometry generated by $R = -i \partial_\theta$. The bosonic DBI$+$CS action on the worldvolume of a single D3 brane is
\begin{equation}
    S_{\mathrm{D3}} = -\frac{N}{2 \pi^2 L^4} \int \left( d^4 x \sqrt{-g_{\mathrm{D3}}} - P[C_4] \right).
\end{equation}
It will be instructive to first work in the stationary frame with respect to $\AdS_5 \times S^5$ to note some properties of the corresponding hamiltonian. The quadratic action $S_{\mathrm{D3}}^{\Phi}$ for transverse fluctuations in $S^5$ in the stationary frame is
\begin{align}
    S_{\mathrm{D3}}^{\Phi} &= \frac{1}{2 \pi^2 L^4} \int dt \ d\Omega_{S^3} \left( |\dot{\Phi}|^2 - |\nabla\Phi|^2 + \frac{3}{ L^2} |\Phi|^2 + \frac{2i}{L} ( \bar{\Phi} \dot{\Phi} - \dot{\bar{\Phi}} \Phi ) \right) \nonumber \\
    &= \frac{1}{2 \pi^2 L^4} \int dt \ d\Omega_{S^3} \left( |\partial_t ( e^{-2 i t/L} \Phi )|^2 - |\nabla\Phi|^2 - \frac{1}{ L^2} |\Phi|^2 \right)
\end{align}
in local complex coordinates $\Phi = \frac{1}{\sqrt{2}} r e^{i \theta}$ and $\bar{\Phi} = \frac{1}{\sqrt{2}} r e^{-i \theta}$. The potential term $- \frac{1}{ L^2} |\Phi|^2$ is present in the second line because of the nonvanishing Ricci scalar on $\mathbb{R} \times S^3$. As was observed in \cite{Arai:2019xmp}, the stationary frame action of a D3 giant is equivalent to the usual action of $\mathcal{N}=4$ SYM on $\mathbb{R} \times S^3$ where its fields possess a time-dependent phase
\begin{equation} \label{eq:phasetwist1}
(\mathrm{field}) \to e^{-\frac{2 i t}{L} R} (\mathrm{field})
\end{equation}
proportional to the relevant worldvolume R-charge, which we can identify with their charge $R$ under the background $U(1)_X$ R-isometry. The scalars  $\Phi, \bar\Phi$ have charge $\pm 1$ under $U(1)_X$.

The resulting worldvolume theory is a $U(1)$ $\mathcal{N}=4$ SYM with a modified potential for fields that carry nonzero $R$. Let us take gauginos with $SU(4)$ indices $\psi^{A=1,2}$ and $\psi^{B=3,4}$ to have R-charges $+\frac{1}{2}$ and $-\frac{1}{2}$, respectively, and let $\mathcal{A}=1,\cdots,4$ be the usual $SU(4)$ index.  Then under \eqref{eq:phasetwist1}, the fermion lagrangian acquires the mass terms
\begin{equation}
+ \frac{1}{L} \bar\psi_A \psi^A - \frac{1}{L} \bar\psi_B \psi^B.
\end{equation}
With these fermion masses, one finds that supersymmetry transformations require conformal Killing spinors $\zeta_\alpha$ on $\mathbb{R} \times S^3$ that depend on time \cite{Festuccia:2011ws}. Therefore, the supercharges depend on time in the stationary frame with respect to the $\AdStimesS$ background, i.e. the generator of time translations in this system is not supersymmetric.

That the generator $H_{\mathrm{D3}}$ of time translations is not supersymmetric has an interesting consequence. In the stationary frame, the hamiltonian is
\begin{align}
    H_{\Dthree} = \int d\Omega_{S^3} \bigg[ &|\Pi|^2 + |\nabla\Phi|^2 + \frac{1}{L^2}|\Phi|^2 + \frac{2 i}{L} ( \Pi \Phi - \bar\Phi \bar\Pi ) \nonumber \\
    &+i \bar\psi_{\mathcal{A}} \bar\sigma^i \nabla_i \psi^{\mathcal{A}} - \frac{1}{L} \bar\psi_A \psi^A + \frac{1}{L} \bar\psi_B \psi^B  \bigg] + \cdots
\end{align}
where ellipses denote contributions from fields on the worldvolume SYM theory that do not possess $U(1)_X$ R-charges. The conjugate momenta $\Pi = \dot{\bar{\Phi}} + \frac{2 i}{L} \bar\Phi$ and $\bar\Pi = \dot{\Phi} - \frac{2 i}{L} \Phi$ and the fermionic counterpart satisfy the equal-time (anti-)commutation relations. Observe that $H_\Dthree$ has the shift $+\frac{2}{L}R$ relative to the free $\mathcal{N}=4$ hamiltonian, where the worldvolume charge $R$ is
\begin{equation}
R = \int d\Omega_{S^3} \left[ i ( \Pi \Phi - \bar\Phi \bar\Pi ) - \frac{1}{2} (\bar\psi_A \psi^A - \bar\psi_B \psi^B)  \right ].
\end{equation}
The hamiltonian $H_\Dthree$ indicates that the energies of D3 worldvolume fields must be shifted by two units of their $R$ charges, relative to naive dimensional analysis based on a $\mathcal{N}=4$ action. That is, the background Ramond-Ramond potential $C_4$ induced a twist of the stress tensor on the worldvolume of a D3 giant. The definition of $R$ here coincides with its previous definition as the background $U(1)_X$ isometry generator $R = -i \partial_\theta$ upon restricting to bosonic $S^3$-zeromodes.

Due to the twist, the only worldvolume field satisfying the half-BPS condition
\begin{equation}
H_{\mathrm{D3}} - \tfrac{1}{L}R = 0
\end{equation}
is the anti-holomorphic scalar $\bar{\Phi}$ with $H_{\mathrm{D3}} L = R = -1$. The negative mode is an artifact of the hamiltonian $H_\Dthree$ not commuting with supercharges in this frame.

\subsection{Localization at small radius} \label{subsec:localization}

In this section, we find the half-BPS open string states on D3 giant gravitons. 

In deriving the worldvolume twist, we considered the D3 action in the stationary frame relative to the $\AdS_5 \times S^5$ background. However, a maximal giant graviton rotates at the speed of light $\dot\theta = \frac{1}{L}$, so expanding about the solution actually entails working in a frame that is rotating with the giant. The worldvolume action about the rotating D3 solution turns out to admit time-independent supercharges for the relevant chiral multiplet, i.e. the time translation generator commutes with these supercharges.

In the worldline quantum mechanics comprising of the lightest $S^3$ modes, these supercharges will be identified with the Dolbeault differential $\bar\partial$ and its adjoint $\bar\partial^\dagger$ deformed by an inverted Morse function $h$. The deformed Dolbeault laplacian $\Delta_h^{\bar\partial} = \{ \bar\partial_h, \bar\partial_h^\dagger \}$ can therefore be identified with the hamiltonian of this supersymmetric quantum mechanics. The BPS states are then the ground states of $\Delta_h^{\bar\partial}$. We find that these states survive and localize to fixed points of R-isometries in the large curvature limit $L \to 0$ of $\AdStimesS$. This allows us to compare the spectrum of the string states with the spectrum of finite $N$ null states in the boundary gauge theory in Section \ref{subsec:ghostsfortracerelations}.

Expanding the action $S_{\mathrm{D3}}^{\Phi}$ around the on-shell value $\theta(t) = \frac{1}{L} t + \delta \theta(t)$, we find
\begin{align}
    S_{\mathrm{D3}}^{\phi} &= \frac{1}{2 \pi^2 L^4} \int dt \ d\Omega_{S^3} \left( |\dot{\phi}|^2 - |\nabla\phi|^2 + \frac{i}{L} ( \bar{\phi} \dot{\phi} - \dot{\bar{\phi}} \phi ) \right) \nonumber \\
    &= \frac{1}{2 \pi^2 L^4} \int dt \ d\Omega_{S^3} \left( |\partial_t ( e^{-i t/L} \phi )|^2 - |\nabla\phi|^2 - \frac{1}{ L^2} |\phi|^2 \right).
\end{align}
The complex coordinates $\phi = \frac{1}{\sqrt{2}} r e^{i \delta\theta}$ and $\bar{\phi} = \frac{1}{\sqrt{2}} r e^{-i \delta\theta}$ are now defined in terms of the fluctuation angle $\delta\theta$. The fields now possess the time-dependent phase
\begin{equation}
(\mathrm{field}) \to e^{-\frac{i t}{L} R} (\mathrm{field})
\end{equation}
compared to those in the usual action of $\mathcal{N}=4$ SYM on $\mathbb{R} \times S^3$. This phase differs from that in the stationary frame by a unit of the R-charge.

We again take gauginos with $SU(4)$ indices $\psi^{A=1,2}$ and $\psi^{B=3,4}$ to have R-charges $+\frac{1}{2}$ and $-\frac{1}{2}$, respectively, and let $\mathcal{A}=1,\cdots,4$ be the usual $SU(4)$ index.  Then the fermion lagrangian acquires the mass terms
$+\frac{1}{2L} \bar\psi_A \psi^A - \frac{1}{2L} \bar\psi_B \psi^B$. The total action becomes
\begin{align} \label{eq:D3 action nonrotating}
    S_{\mathrm{D3}} = \frac{1}{2 \pi^2 L^4} \int dt \ d\Omega_{S^3} &\bigg[ -\partial_\mu \bar\phi \partial^\mu \phi + \frac{i}{L} ( \bar{\phi} \dot{\phi} - \dot{\bar{\phi}} \phi ) \nonumber \\
    &-i \bar\psi_{\mathcal{A}} \bar\sigma^\mu \nabla_\mu \psi^{\mathcal{A}} + \frac{1}{2L} \bar\psi_A \psi^A - \frac{1}{2L} \bar\psi_B \psi^B  \bigg] + \cdots 
\end{align}
where ellipses denote fields that do not possess $U(1)_X$ R-charges. We will restrict attention to half of the supersymmetries
\begin{align} \label{eq:susytranform half}
    \delta \phi &= -\sqrt{2} \zeta_A \psi^A \nonumber \\
    \delta \bar{\phi} &= -\sqrt{2} \bar\zeta^A \bar\psi_A \nonumber \\
    \delta \psi^A_\alpha &= - i \sqrt{2} (\sigma^\mu \bar{\zeta}^A)_{\alpha} \partial_\mu \phi \nonumber \\
    \delta \bar{\psi}_{A \dot\alpha} &= + i \sqrt{2} (\zeta_A \sigma^\mu)_{\dot\alpha} \partial_\mu \bar\phi
\end{align}
that relate the complex scalars $\phi,\bar\phi$ for transverse fluctuations of $S^3$ in $S^5$ with only half of the gauginos $\psi^{A}, \bar\psi_{A}$ of positive R-charge. The corresponding conformal Killing spinors $\zeta,\bar\zeta$ on $\mathbb{R} \times S^3$ satisfy \cite{Festuccia:2011ws}
\begin{align}
    0 &= \partial_t \zeta_A^\alpha \nonumber \\
    0 &= \partial_t \bar\zeta^A_{\dot\alpha} \nonumber \\
    0 &= \nabla_i \zeta_A^\alpha + \frac{i}{L} (\zeta_A \sigma_{0i})^\alpha \nonumber \\
    0 &= \nabla_i \bar\zeta^A_{\dot\alpha} - \frac{i}{L} (\bar\zeta^A \bar\sigma_{0i})_{\dot\alpha}.
\end{align}
The spinors $\zeta,\bar\zeta$ are time-independent unlike the previous case, so the hamiltonian of this worldvolume theory commutes with the supercharges from \eqref{eq:susytranform half}.

The new generator $H_{\mathrm{D3}}'$ of time translations for the D3 solution
\begin{align} \label{eq:newhamiltonian}
    H_{\Dthree}' = \int d\Omega_{S^3} \bigg[ &|\pi|^2 + |\nabla\pi|^2 + \frac{i}{L} ( \pi \phi - \bar\phi \bar\pi ) \nonumber \\
    &+i \bar\psi_{\mathcal{A}} \bar\sigma^i \nabla_i \psi^{\mathcal{A}} - \frac{1}{2L} \bar\psi_A \psi^A + \frac{1}{2L} \bar\psi_B \psi^B  \bigg] + \cdots
\end{align}
has the shift $+\frac{1}{L}R$ relative to the free $\Nfour$ hamiltonian, where the definition of the R-charge
\begin{equation}
R = \int d\Omega_{S^3} \left[ i ( \pi \phi - \bar\phi \bar\pi ) - \frac{1}{2} (\bar\psi_A \psi^A - \bar\psi_B \psi^B)  \right ].
\end{equation}
remains invariant. The conjugate momenta $\pi = \dot{\bar{\phi}} + \frac{i}{L} \bar\phi$ and $\bar\pi = \dot{\phi} - \frac{i}{L} \phi$ and the fermionic counterpart satisfy the equal-time (anti-)commutation relations. $H_\Dthree'$ has the shift $-\frac{1}{L}R$ relative to the stationary frame hamiltonian $H_\Dthree$ upon relating the fields and momenta, so we recognize the new hamiltonian
\begin{equation}
H_\Dthree' = H_\Dthree - \tfrac{1}{L} R
\end{equation}
as the supersymmetric hamiltonian whose ground states are BPS states.

\subsubsection{Reduction to quantum mechanics} \label{subsubsec:quantummechanics}

For the half-BPS sector, it actually suffices to consider the supersymmetric quantum mechanics given by consistent truncation to the lightest $S^3$-modes related by \eqref{eq:susytranform half}, because the higher modes do not satisfy the BPS condition. See e.g. \cite{Ishiki:2006rt} for details on the $S^3$ mode expansion of $\Nfour$ SYM on $\mathbb{R} \times S^3$, though in our case the Killing spinors $\zeta, \bar\zeta$ are time-independent. The fluctuation scalars and gauginos have $S^3$ mode expansions
\begin{align}
    \phi(t,\Omega) &= \sum_{j=0}^\infty \sum_{M} \phi_{j M}(t) Y_{j M}(\Omega) \nonumber \\
    \psi^{\mathcal{A}}_\alpha(t,\Omega) &= \sum_{j=0}^\infty \sum_{\kappa = \pm} \sum_{M} \psi_{j M \kappa}^{\mathcal{A}}(t) Y_{j M \alpha}^{\kappa}(\Omega).
\end{align}
$j$ labels the irreps of the $S^3$ isometry group $SO(4) \simeq SU(2)_L \times SU(2)_R$ , which are $(\frac{j}{2},\frac{j}{2})$ for scalars and $(\frac{j+1}{2},\frac{j}{2})$, $(\frac{j}{2},\frac{j+1}{2})$ for the $\kappa = \pm$ fermions, respectively. $M$ runs over the dimensions of the irreducible representation associated with $j$. We will be interested in the scalar $S^3$ zeromode $\phi \equiv \phi_{00}(t)$ and the lightest gaugino mode $\psi \equiv \psi^{1}_{0M+}(t)$ with $R=+\frac{1}{2}$, $\kappa=+$, and $M=(+\frac{1}{2},0)$.\footnote{A choice of fermion modes with $R = -\frac{1}{2}$ and $\kappa=-$ would correspond to a different set of supersymmetries.} The other choice $M=(-\frac{1}{2},0)$ or linear combinations thereof are equally valid. We may also include higher modes to generalize our computation to BPS sectors preserving fewer supersymmetries, but we leave this to future work. 

The worldline action for half-BPS excitations of a D3 giant is
\begin{equation}
    S_{\mathrm{QM}} = \frac{N}{L} \int dt \bigg[ \dot{\bar\phi} \dot\phi + \frac{i}{L} \left( \bar\phi \dot\phi - \dot{\bar\phi} \phi \right) + i \bar\psi \dot{\psi} + \frac{1}{L} (\bar\psi \psi - \psi \bar\psi) \bigg].
\end{equation}
While this action is invariant under simple supersymmetries that descend from \eqref{eq:susytranform half}, the shift in the definition of conjugate momenta results in an interesting interpretation of the supercharges on the transverse plane to $S^3$ in $S^5$, which is locally $\mathbb{C}$.\footnote{The $S^3$-zeromode conjugate momenta are $\pi = \dot{\bar{\phi}} + \frac{i}{L} \bar\phi$ and $\bar\pi = \dot{\phi} - \frac{i}{L} \phi$, and the (anti-)commutation relations
\begin{equation}
[\phi,\pi] = i, \qquad [\bar\phi,\bar\pi] = i, \qquad \{ \psi,\bar\psi \} = 1
\end{equation}
hold at the level of the relevant $S^3$ modes.} In particular, we can identify the supercharges
\begin{align}
    Q &= i \bar\psi \dot{\bar\phi} = \bar\psi ( i \bar\pi - \tfrac{1}{L} \phi ) \nonumber \\
    \bar{Q} &= -i \psi \dot\phi = \psi (-i \pi - \tfrac{1}{L} \bar\phi )
\end{align}
as the Dolbeault differential and its adjoint\footnote{The adjoint of the Dolbeault differential $\bar\partial$ in one complex dimension is defined as $\bar\partial^\dag = - \star \partial \star$.}
\begin{align}
    \bar\partial_h &= e^{-h/L} \, \bar\partial \, e^{h/L} \nonumber \\
    \bar\partial_h^\dag &= e^{h/L} \, \bar\partial^\dag \, e^{-h/L}
\end{align}
deformed by an \textit{inverted} Morse function $h = -\bar\phi \phi$. The inverse radius $L^{-1}$ plays the role of the localization parameter. Then the hamiltonian $H = \{ Q, \bar Q \}$ of this supersymmetric quantum mechanics
\begin{equation}
H = \bar\pi \pi + \frac{1}{L^2} \bar\phi \phi + \frac{i}{L} (\phi \pi - \bar\phi \bar\pi) - \frac{1}{L} [ \bar\psi, \psi]
\end{equation}
is the deformed Dolbeault laplacian
\begin{equation}
\Delta_h^{\bar\partial} = \bar\partial_h \bar\partial_h^\dag + \bar\partial_h^\dag \bar\partial_h.
\end{equation}
We also identify the fermions with operations $\bar\psi = d\bar\phi \ \wedge$ and $\psi = i_{\partial/\partial \bar\phi}$ on differential forms. The BPS states are the eigenstates of $H = \Delta_h^{\bar\partial}$ with zero eigenvalue. This system was studied in the context of holomorphic Morse inequalities in \cite{Witten:1984hol}.

The function $h$ induces a vector field $V$ generating an isometry of $\mathbb{C}$ with fixed points at the critical points of $h$. Near any fixed point, the Dolbeault laplacian deformed by $h$ takes the form
\begin{equation} \label{eq:dolbeault laplacian deformed}
\Delta_h^{\bar\partial} = \Delta^{\bar\partial} + \frac{1}{L^2} \bar\phi \phi - \frac{1}{L} \mathcal{L}_V + \frac{1}{L} (\cdots) 
\end{equation}
where $\mathcal{L}_V$ is the Lie derivative with respect to the vector field $V$. The ellipses denote scalar terms that do not contain derivatives. For $H$, the induced vector field $V$ is
\begin{equation}
V = - \left( \phi \frac{\partial}{\partial \phi} - \bar\phi \frac{\partial}{\partial \bar \phi} \right),
\end{equation}
which coincides with the $U(1)_X$ isometry generator up to a sign. The sign determines the types of states that have support at the fixed point $p$ of $V$ as well as the orientation for the isometry action $\mathcal{L}_V$ that is appropriate for computing the R-charges of these states. We elaborate on this point shortly.

Let us discuss why it suffices to consider only the quadratic fluctuation action of the maximal D3 solution for the purpose of finding BPS states that survive in the small radius limit of $\AdStimesS$. Taking this limit is within the regime of validity of the worldline quantum mechanics on $\mathbb{R}$, because we take this limit after having decomposed fields into $S^3$ harmonics. We also neglect the higher open string fields because they do not satisfy the BPS condition. First we note that since $V$ generates an isometry of the $\phi$-plane, we have
\begin{equation}
[\Delta_h^{\bar\partial} , \mathcal{L}_V] = 0.
\end{equation}
Therefore we can find the relevant BPS states by finding solutions to
\begin{equation}
\Delta_h^{\bar\partial} \Psi = 0
\end{equation}
in the small radius limit $L \to 0$, within each $\mathcal{L}_V$-eigenspace
\begin{equation}
\mathcal{L}_V \Psi = m \Psi
\end{equation}
of fixed charge $m$. Then the dominant potential $\frac{1}{L^2} \bar\phi \phi$ forces all BPS states of charge $m$ to be localized near the fixed point, for any given $m$ in the limit $L \to 0$. That is, BPS open string states of D3 giants localize to fixed points of bulk isometries in the small radius limit $L \to 0$ of $\AdStimesS$. The quadratic D3 action was sufficient because $\frac{1}{L^2} \bar\phi \phi$ is the dominant potential at any zero of $V$. 

Our localization argument explains why dual D3 giant gravitons expanding in $S^3 \subset \AdS_5$ do not seem to appear in the giant graviton expansion. A dual giant with large R-charge does not sit at a fixed point of the R-isometry, so BPS states of dual giants do not survive the small radius limit.

\subsubsection{Normalizable states} \label{subsubsec:normalizablestates}

To find normalizable BPS states, we allow $\Delta_h^{\bar\partial}$ to act on the space of differential forms.

Recall that the Morse index (for real manifolds) of a critical point $p$ is the number of unstable directions at $p$ with respect to the function $h$. The holomorphic analog $n_p$ of the Morse index \cite{Witten:1984hol} is the number of unstable complex directions at the critical point $p$ (with respect to $h$), or equivalently the number of complex directions for which $V$ acts clockwise in local complex coordinates about the fixed point $p$. The holomorphic Morse index $n_p$ coincides with the anti-holomorphic form degree of the normalizable state supported near $p$.

As in \cite{Witten:1984hol}, the Lie derivative $\mathcal{L}_V$ acting on forms with respect to $V$ computes the R-charge of a BPS state at $p$, given a choice of the fugacity domain for which the charge spectrum converges. In particular, the operator $\mathcal{L}_V$ computes directly the charge spectrum of states that is convergent in the fugacity domain by taking into account the orientation at $p$ induced by $h$. For example, for a fixed point with $n_p=1$, the normalizable states are anti-holomorphic 1-forms so $\mathcal{L}_V$ comptues the R-charge with a reversed orientation for the isometry action. 

For the problem at hand, $p$ is the sole fixed point of $V$ at the origin of the $\phi$-plane and it has holomorphic Morse index $n_p = 1$. The normalizable tower of BPS states is given by anti-holomorphic 1-forms
\begin{equation}
\Psi_m = \bar\phi^m e^{-\bar{\phi}\phi  / L} d\bar{\phi}
\end{equation}
that satisfy
\begin{equation}
\bar\partial_h^\dag \Psi_m = \left( - \frac{\partial}{\partial \phi} - \frac{1}{L} \bar\phi \right) \Psi_m = 0
\end{equation}
or $\Delta_h^{\bar\partial} \Psi_m = 0$, for $m=0,1,2, \cdots$. In the half-BPS sector we are interested in the fugacity domain $|x|<1$ where each BPS state is counted with weight $x^R$. We can directly compute the charge spectrum that is convergent for $|x|<1$:
\begin{equation}
\mathcal{L}_V \Psi_m = (m+1) \Psi_m.
\end{equation}
This tower gives the half-BPS open string spectrum on a single giant
\begin{equation} \label{eq:single giant spectrum}
\mu \ x^N \sum_{m=0}^\infty x^{m+1} = \mu \, x^N \frac{x}{1-x},
\end{equation}
where $x^N$ accounts for R-charges carried by the D3 giant and we introduced a fugacity $\mu$ for the form degree. This coincides with the $k=1$ term of the giant graviton expansion in the half-BPS sector with $\mu = -1$.

There is a shift
\begin{equation}
N \to N+1
\end{equation}
in the R-charge of the open string vacuum relative to the classical expectation of $N$. This resulted from the fact that the open string vacuum on a single D3 giant is labelled by an anti-holomorphic 1-form.

Note that a naive count of the charges based on the background R-isometry (and without accounting for form-valued states) would have given the spectrum $x^N \frac{1}{1-x^{-1}}$, whose series expression converges for $|x|>1$. This series is related to \eqref{eq:single giant spectrum} by analytic continuation.

\subsubsection{Generalization to $k$ coincident giants} \label{subsubsec:kgiants}

Let us generalize the analysis to $k$ coincident giants. We could diagonalize the hamiltonian
\begin{align}
    H_k &= \Tr \left[ \bar\pi \pi + \frac{1}{L^2} \bar\phi \phi + \frac{i}{L} (\phi \pi - \bar\phi \bar\pi) - \frac{1}{L} [ \bar\psi, \psi] \right] \nonumber \\
    &= \Tr \left[ - \frac{\partial}{\partial \bar\phi} \frac{\partial}{\partial \phi} + \frac{1}{L^2} \bar\phi \phi + \frac{1}{L} \left( \phi \frac{\partial}{\partial \phi} - \bar\phi \frac{\partial}{\partial \bar\phi} \right) - \frac{1}{L} ( \bar\psi \psi - \psi \bar\psi ) \right]
\end{align}
in the position representation where the operators are now $k \times k$ matrices, and then look for normalizable $U(k)$-invariant ground states. We proceed instead via a shortcut, namely by finding form-valued and $U(k)$-invariant solutions to the supersymmetry conditions
\begin{equation}
\bar\partial_h \Psi = \bar\partial_h^\dag \Psi = 0
\end{equation}
under a suitable ansatz for the ground state wavefunction $\Psi$.

Let us first determine the local form of the Morse potential $h$ in terms of the eigenvalues $\phi_a, \bar\phi_b$ of the $k \times k$ matrices $\phi, \bar\phi$. Let $P$ and $Q$ be invertible matrices such that $\phi_{ij} = P_{ia} \phi_a P^{-1}_{aj}$ and $\bar\phi_{ij} = Q_{ia} \bar\phi_a Q^{-1}_{aj}$. We have $Q = P^\dag$ since $\phi$ and $\bar\phi$ matrices are hermitian conjugates. The relevant part of the hamiltonian $H_k$ containing the induced vector field $V$ is
\begin{equation}
\frac{1}{L} \Tr \left[ \phi \frac{\partial}{\partial \phi} - \bar\phi \frac{\partial}{\partial \bar\phi} \right] = \frac{1}{L} \sum_{i,j=1}^{k} \left( \phi_{ij} \frac{\partial}{\partial \phi_{ij}} - \bar\phi_{ij} \frac{\partial}{\partial \bar\phi_{ij}} \right).
\end{equation}
We can express the derivatives with respect to matrices $\phi, \bar\phi$ in terms of the eigenvalues
\begin{equation}
\frac{\partial}{\partial \phi_{ij}} = \sum_{a=1}^k P^{-1}_{ai} P_{ja} \frac{\partial}{\partial \phi_a}, \qquad \frac{\partial}{\partial \bar\phi_{ij}} = \sum_{a=1}^k Q^{-1}_{ai} Q_{ja} \frac{\partial}{\partial \bar\phi_a},
\end{equation}
so we have
\begin{align}
    \frac{1}{L} \Tr \left[ \phi \frac{\partial}{\partial \phi} - \bar\phi \frac{\partial}{\partial \bar\phi} \right] &= \frac{1}{L} \sum_{i,j=1}^{k} \sum_{a,b=1}^{k} \left[ P_{ia} \phi_a P^{-1}_{aj} P^{-1}_{bi} P_{jb} \frac{\partial}{\partial \phi_b} - Q_{ia} \bar\phi_a Q^{-1}_{aj} Q^{-1}_{bi} Q_{jb} \frac{\partial}{\partial \bar\phi_b} \right] \nonumber \\
    &= + \sum_{a=1}^k \left( \phi_a \frac{\partial}{\partial \phi_a} - \bar\phi_b \frac{\partial}{\partial \bar\phi_b} \right).
\end{align}
The local form of the vector field $V$ about the origin in the space $\mathbb{C}^k$ of eigenvalues, found by comparing the above expression with the Lie derivative $-\frac{1}{L}\mathcal{L}_V$ of $\Delta_h^{\bar\partial}$ in \eqref{eq:dolbeault laplacian deformed}, is
\begin{equation}
V = - \sum_{a = 1}^k \left( \phi_a \frac{\partial}{\partial \phi_a} - \bar\phi_b \frac{\partial}{\partial \bar\phi_b} \right).
\end{equation}
$V$ is induced from the Morse function $h = - \sum_{a=1}^k \bar\phi_a \phi_a$ in local coordinates. The holomorphic Morse index at this point is $n_p = k$, so the state $\Psi$ is an anti-holomorphic $k$-form. Thus, the normalizable state $\Psi$ must have the expression
\begin{equation}
\Psi = G(\phi_a, \bar\phi_b) e^{-\sum_{a=1}^k \bar\phi_a \phi_a / L} d\bar\phi_1 \wedge \cdots \wedge d\bar\phi_k
\end{equation}
in local eigenvalue coordinates $(\phi_a, \bar\phi_b)$. $G(\phi_a, \bar\phi_b)$ is totally antisymmetric under exchanges of the anti-holomorphic eigenvalues $\bar\phi_b$.

The supersymmetry condition $\bar\partial_h^\dag \Psi = 0$ imposes further restrictions. Since
\begin{equation}
\bar\partial_h^\dag \Psi \propto \sum_{a=1}^k \frac{\partial G}{\partial \phi_a} d\bar\phi_1 \wedge \cdots \wedge \widehat{d\bar\phi_a} \wedge \cdots \wedge d\bar\phi_k =0,
\end{equation}
where we omit the hatted component, $G$ is independent of the holomorphic eigenvalues $\phi_a$. As there are no further constraints, we can now write a $U(k)$-invariant basis for the supersymmetric ground states of $H_k$:
\begin{equation}
\Psi_{m_1 \cdots m_k} = \frac{1}{k!} \sum_{a_1,\cdots, a_k=1}^k \epsilon_{a_1 \cdots a_k} \bar\phi_1^{m_{a_1}} \bar\phi_2^{m_{a_2}} \cdots \bar\phi_k^{m_{a_k}}  e^{-\sum_{a=1}^k \bar{\phi}_a \phi_a  / L} d\bar\phi_1 \wedge \cdots \wedge d\bar\phi_k
\end{equation}
where $0 \leq m_1<m_2< \cdots<m_k<\infty$. These states have R-charges
\begin{equation}
\mathcal{L}_V \Psi_{m_1 \cdots m_k} = (m_1 + \cdots + m_k + k) \Psi_{m_1 \cdots m_k},
\end{equation}
so the half-BPS spectrum of $k$ giants is
\begin{equation} \label{eq:k giant spectrum}
\mu^k x^{k N} \sum_{0 \leq m_1<\cdots<m_k<\infty} x^{m_1 + \cdots + m_k + k} = \mu^k x^{k N} \frac{x^{k(k+1)/2}}{\prod_{n=1}^k (1-x^n)}
\end{equation}
This is the $k$-th term of the giant graviton expansion in the half-BPS sector with $\mu = -1$.

There is again a shift
\begin{equation}
k N \to k N + \frac{1}{2} k(k+1)
\end{equation}
in the R-charge of the open string vacuum relative to the classical expectation of $k N$, which originates from the fact that the open string vacuum on $k$ D3 giants is labelled by an anti-holomorphic $k$-form.

\subsection{Ghosts for trace relations} \label{subsec:ghostsfortracerelations}

In our bulk analysis, we have seen that the open string vacuum on $k$ giant graviton branes is labelled by a differential $k$-form. The form degree of states resulted from the fact that the supersymmetry generators were identified to be Dolbeault differentials deformed by a Morse function $h$ with $k$ unstable complex directions at the fixed point of an isometry. The form degree equips the bulk open string Hilbert space with an extra grading associated to the number of branes in the background.

On the other hand, normalizable BPS states of the boundary $\Nfour$ super Yang-Mills on $\mathbb{R} \times S^3$ are scalar-valued because the corresponding Morse function does not admit any unstable direction. This essentially stems from the fact that the BPS hamiltonian $H_\BPS = H_\Nfour - R$ of the boundary gauge theory possesses an angular momentum term that is opposite in sign compared to that of $H_\Dthree'$ in \eqref{eq:newhamiltonian}. An analysis similar to that done in Section \ref{subsubsec:kgiants} shows that the following normalizable wavefunctions
\begin{equation}
\Psi_{m_1 \cdots m_N} = \frac{1}{N!} \sum_{\sigma \in S_N} X_1^{m_{\sigma(1)}} X_2^{m_{\sigma(2)}} \cdots X_N^{m_{\sigma(N)}}  e^{-\sum_{a=1}^N \bar{X}_a X_a},
\end{equation}
with $0 \leq m_1 \leq m_2 \leq \cdots \leq m_N <\infty$, form a basis of half-BPS states for the boundary $U(N)$ $\Nfour$ SYM. These states were written in terms of the eigenvalues $(X_a, \bar{X}_b)$ of the $N \times N$ scalars $X$ and $\bar{X}$. They have R-charges
\begin{equation}
\mathcal{L}_V \Psi_{m_1 \cdots m_N} = (m_1 + \cdots + m_N) \Psi_{m_1 \cdots m_N}
\end{equation}
with $V = + \sum_{a=1}^N \left( X_a \frac{\partial}{\partial X_a} - \bar{X}_a \frac{\partial}{\partial \bar{X}_a} \right)$, and their counting reproduces the boundary half-BPS spectrum $Z_N = \prod_{n=1}^N \frac{1}{1 - x^n}$. This shows that the grading with respect to the form degree of states does not appear in the physical Hilbert space of the boundary gauge theory.

The obvious question is then, how can the presence of the extra grading in the bulk open string Hilbert space be consistent with holography? To answer this question, we need to identify the ``states'' in the boundary gauge theory that possess an extra grading and also reproduce the spectrum \eqref{eq:k giant spectrum}. Identifying such states is made subtle because, as demonstrated above, these states cannot belong to the physical Hilbert space of the gauge theory.

We can obtain clues regarding the nature of these states by examining the ``mode'' expansion in $z$ of the determinant operator $\det (z - X)$ inserted at a point in $U(N)$ $\mathcal{N}=4$ SYM. This operator is dual to a family of D3 giant graviton branes wrapped on $S^3 \subset S^5$ with order $N$ units of an R-charge. The auxiliary parameter $z$ has an interpretation as the bulk insertion point of the brane on the plane transverse to the $S^3$ in $S^5$. More explicitly, according to $\det (z - X)$ the brane must reach its maximal size at $z=0$ and shrink to zero size at $z \to \infty$. We can therefore make the identification $z = \frac{\Phi}{1 - \Phi \bar{\Phi}/L^2}$ in coordinates discussed in Section \ref{subsec:twistworldvolume}. The spacetime interpretation of $z$ is useful even though we are working in a regime where stringy effects are large.

The fluctuation ``modes'' in $z$ of the determinant operator are
\begin{equation}
\det ( z - X) = z^N e^{-\sum_{n=1}^\infty\frac{1}{n} z^{-n} \Tr X^n} = z^N \sum_{n=0}^\infty \frac{(-1)^n P_{n}}{z^{n}}.
\end{equation}
where $P_n = P_n ( \Tr X^{\bullet} )$ are defined in terms of the exponential generating function and where the expression is normal-ordered. The first few modes are
\begin{align}
    P_1 &= \Tr X \nonumber \\
    P_2 &= \frac{1}{2} \left( (\Tr X)^2 - (\Tr X^2) \right) \nonumber \\
    P_3 &= \frac{1}{6} \left( (\Tr X)^3 - 3 (\Tr X) (\Tr X^2) + 2 (\Tr X^3) \right) \nonumber \\
    P_4 &= \frac{1}{24} \left( (\Tr X)^4 - 6 (\Tr X)^2 (\Tr X^2) + 3 (\Tr X^2)^2 + 8 (\Tr X) (\Tr X^3)  - 6 (\Tr X^4) \right)
\end{align}
and so on. At any positive integer value $N$, the tower of modes
\begin{equation}
\det ( z - X) = (-1)^{N+1} \left[ \cdots + \frac{P_{N+1}}{z} - \frac{P_{N+2}}{z^2} + \frac{P_{N+3}}{z^3} + \cdots \right]
\end{equation}
starting at $P_{N+1}$ are but the trace relations between gauge-invariant operators that are present in any $U(N)$ gauge theory. In other words, states constructed by acting with modes $P_{N+1}, P_{N+2}, \cdots$ on the vacuum $|0 \rangle$ are the finite $N$ null states. Since these modes vanish automatically at any integer $N$, the determinant operator $\det (z - X)$ is a polynomial of degree $N$ in $z$.

While these null states yield the correct half-BPS spectrum $x^N \frac{x}{1-x}$ of a single D3 giant, we should not identify the set of trace relations directly with the states of bulk D-branes. The reason is that one needs the states to possess an extra grading in order to explain the overall signs $(-1)^k$ that appear in the BPS spectrum on $k$ coincident D-branes, and no such grading distinguishes the finite $N$ null states from the physical states.

Instead, we identify the BPS D-brane states with auxiliary ``heavy'' ghosts $\chi$, with dimension and R-charge starting at $N+1$, in a boundary $U(\infty)$ gauge theory that compensate for null states due to trace relations at a finite value of $N$. The effect of introducing these anti-commuting ghosts along with a homological charge $\hQ$ acting as
\begin{align}
    [\chi_{-N-a},\hQ]_\pm &= P_{N+a} \nonumber \\
    [P_a, \hQ]_\pm &= 0 ,
\end{align}
where $a=1,2, \cdots$, is to reduce the $U(\infty)$ theory to a $U(N)$ gauge theory. We assign the ``heavy'' ghost number $1$ to $\chi_a$ and $-1$ to the differential $\hcQ = [\, \cdot \, ,\hQ]_\pm$. The heavy ghosts should be distinguished from the BRST (anti-)ghosts that one introduces for gauge-fixing and to impose on-shell constraints in a $U(N)$ gauge theory. Rather, the role of the heavy ghost and the differential is to supplement the $U(\infty)$ theory with directions for systematically removing all but the $U(N)$ degrees of freedom.

It is simple to compute the spectrum with ghost number $k$ in the half-BPS sector, because this sector does not have ghosts-for-ghosts. The states with ghost number $k$ are therefore multi-ghost states built out of $\chi_{-N-a}$. We can write a basis of $k$ multi-ghost states as
\begin{equation}
\chi_{-N-a_1} \chi_{-N-a_2} \cdots \chi_{-N-a_k} |0\rangle
\end{equation}
where $1 \leq a_1 < a_2 < \cdots < a_k < \infty$, which yields the charge spectrum
\begin{equation}
\mu^k x^{k N} \frac{x^{k(k+1)/2}}{\prod_{n=1}^k (1- x^n)}
\end{equation}
that agrees with our bulk computation \eqref{eq:k giant spectrum}. Fugacity $\mu$ in the gauge theory now keeps track of the heavy ghost number.

Our findings suggest that the form degree $k$ for the states of $k$ D-branes in the bulk should be identified with the ghost number assigned to auxiliary ghosts that compensate for trace relations in the boundary gauge theory (as opposed to the fermion number $F$ in the gauge theory that is usually defined in terms of other quantum numbers). This is the extra grading in the bulk that is not seen in the physical Hilbert space of a $U(N)$ gauge theory if one works strictly at any positive integer value of $N$.

\section{Bulk Hilbert space as a resolution of finite $N$ modules} \label{sec:bulkhilbertspace}

Thus far, we found the BPS states of coincident bulk D-branes by quantizing its open string modes. These states possess an additional grading, associated with the brane number $k$ labelling the open string vacua, whose bulk origin is the form degree of states and whose boundary origin is the ghost number assigned to ``heavy'' ghosts that compensate for finite $N$ null states.

We have not yet discussed how the space of states built on different D-brane backgrounds are related to each other. That is, we should ask whether there is a natural way to organize the relationships among the perturbative Hilbert spaces built on various D-brane backgrounds.

In this section, we comment on global aspects of the bulk Hilbert space at vanishing 't Hooft coupling $\lambda$ and finite $N$, by employing the holographic map between D-brane states and heavy ghosts. Our proposal is that a complex formed from the spaces of states of certain bulk D-branes in the $\alpha' \to \infty$ limit furnishes a BRST-like resolution of the finite $N$ Hilbert space of the boundary gauge theory at $\lambda = 0$. Objects in the complex are labelled by the heavy ghost number $k$. The giant graviton expansion can be derived by computing the Hilbert series of the resolution.

As discussed in Section \ref{subsubsec:quantummechanics}, taking the small radius limit $L \to 0$ of D-brane states is well-defined if they are the ground states of a supersymmetric hamiltonian of the worldline quantum mechanics on $\mathbb{R}$. Such states localize to fixed points of bulk isometries at large curvatures because the inverse radius $L^{-1}$ assumes the role of a localization parameter. The consideration of D-branes at finite $N$ is, of course, an extrapolation drawn from our computations at large $N$. That the extrapolation from large to finite $N$ works in the giant graviton expansion is rather miraculous, and we discuss its implications in Section \ref{subsubsec:largeN}.

While the bulk computations of Section \ref{sec:dbraneghost} relied on supersymmetry and the presence of a weakly-curved gravity limit, there are reasons to think that the identification between D-brane states and heavy ghosts is quite general and holds independently of those properties. Trace relations between gauge-invariant operators are present in any finite $N$ gauge theory with adjoint fields, and such a theory can be rewritten as a $N=\infty$ gauge theory supplemented with an auxiliary ghost for every trace relation in the original theory.\footnote{There can certainly be non-trivial relations between trace relations that necessitate ghosts-for-ghosts, and we discuss these cases thoroughly.}

\subsection{Half-BPS Hilbert space in the bulk} \label{subsec:halfbpsbulk}

Here, we motivate our main proposal regarding the bulk Hilbert space by gathering the lessons of our bulk computations in the half-BPS sector. The half-BPS sector is rather special because the set of trace relations do not satisfy further non-trivial relations. Therefore, one does not need to introduce ghosts-for-ghosts. We generalize these lessons subsequently.

In Section \ref{subsec:localization}, we found the states of D-branes responsible for the spectrum
\begin{equation}
\mu^k x^{k N} \frac{x^{k(k+1)/2}}{\prod_{m=1}^k (1- x^m)}
\end{equation}
on $k$ giant graviton branes. The giant graviton expansion suggests that, to retrieve the half-BPS spectrum of boundary $\Nfour$ SYM at finite $N$, we should simply sum up the open string spectra $\hat{Z}_k(x)$ of $k$-brane sectors for all $k$, each in the presence of half-BPS closed strings $Z_\infty(x)$ of the $\AdStimesS$ background:
\begin{equation} \label{eq:halfBPS gge bulk}
\frac{1}{\prod_{n=1}^N (1-x^n)} = \frac{1}{\prod_{n=1}^\infty (1-x^n)} \sum_{k=0}^\infty (-1)^k x^{k N} \frac{x^{k(k+1)/2}}{\prod_{m=1}^k (1 - x^m)}.
\end{equation}
We also set the fugacity $\mu$ for the form degree grading to $\mu = -1$.

On the other hand, we identified its holographic dual as heavy ghosts $\chi$ that compensate for the presence of finite $N$ trace relations. We can study the half-BPS sector at finite $N$ by equipping the $U(\infty)$ gauge theory with ghosts $\chi$ and a homological differential $\hcQ = [\, \cdot \, , \hQ]_\pm$ acting on operators via the graded Leibniz rule
\begin{align}
    \hcQ ( \chi_{-N-a_1} \cdots \chi_{-N-a_k}) &= \sum_{i=1}^{k} (-1)^{i+1} ( \chi_{-N-a_1} \cdots \hcQ(\chi_{-N-a_i}) \cdots \chi_{-N-a_k} ) \nonumber \\
    &= \sum_{i=1}^{k} (-1)^{i+1} P_{N+a_i} ( \chi_{-N-a_1} \cdots \chi_{-N-a_{i-1}} \chi_{-N-a_{i+1}} \cdots \chi_{-N-a_k} )
\end{align}
and $\hcQ^2 = 0$. In particular, the ghosts $\chi$ map to trace relations: $\hcQ (\chi_{-N-a}) = P_{N+a}$. Then the space of multi-ghost states, with coefficients in the ring of polynomials of traces
\begin{equation}
R = \mathbb{C}[\Tr X, \Tr X^2, \Tr X^3, \cdots ],
\end{equation}
is given by the complex
\begin{equation} \label{eq:halfBPS truncated complex}
\cdots \ \rightarrow \ \cV_3 \ \xrightarrow{\hQ} \ \cV_2 \ \xrightarrow{\hQ} \ \cV_1 \ \xrightarrow{\hQ} \ \cH_\infty \ \rightarrow \ 0,
\end{equation}
where $\cV_k$ is the space of multi-ghost states of heavy ghost number $k$ with coefficients in the ring $R$. Also, $\cV_0 \equiv \cH_\infty = R | 0 \rangle$ is the half-BPS Hilbert space of the gauge theory at infinite $N$. The ghost degree $k$ of the complex labels the open string vacuum associated to $k$ coincident D3 branes in the bulk.

The complex \eqref{eq:halfBPS truncated complex} has non-vanishing homology only at zero ghost number. Since the image $\hQ \cV_1 \subset \cH_\infty$ is the space of trace relations with coefficients in $R$, the homology at ghost number $k=0$ is the Hilbert space
\begin{equation}
\cH_N = \cH_\infty / \hQ \cV_1.
\end{equation}
in the half-BPS sector of $U(N)$ $\Nfour$ SYM at finite $N$.

Let us observe that each $\cV_k$, $k=0,1,2, \cdots$ in the complex \eqref{eq:halfBPS truncated complex} is freely generated by the multi-ghost generators
\begin{equation}
\chi_{-N-a_1} \chi_{-N-a_2} \cdots \chi_{-N-a_k} |0\rangle
\end{equation}
with coefficients in $R$. In other words, $\cV_k$ are free $R$-modules. This means that, if we augment \eqref{eq:halfBPS truncated complex} with $\cH_N$ at the final step, the resulting exact sequence
\begin{equation} \label{eq:halfBPS resolution}
\cdots \ \rightarrow \ \cV_3 \ \xrightarrow{\hQ} \ \cV_2 \ \xrightarrow{\hQ} \ \cV_1 \ \xrightarrow{\hQ} \ \cH_\infty \rightarrow \cH_N \rightarrow 0
\end{equation}
is a free resolution of the half-BPS Hilbert space $\cH_N$ at finite $N$. We will explain these notions in Section \ref{subsec:mathematicalbackground}. For now, it suffices to realize that the resolution \eqref{eq:halfBPS resolution} gives a precise relationship between (1) the half-BPS Hilbert space $\cH_N$ of $U(N)$ $\Nfour$ SYM at finite $N$ and (2) the complex \eqref{eq:halfBPS truncated complex} identified as the ``bulk'' half-BPS Hilbert space of IIB string theory on $\AdStimesS$ via the holographic map between D-brane states and heavy ghosts. The resolution says that the open strings on bulk D-branes are encoded in the \textit{relations} between the single-trace generators that arise in the boundary gauge theory at finite $N$.

It is simple to use this observation to derive the giant graviton expansion \eqref{eq:halfBPS gge bulk} in the half-BPS sector. Since an alternating sum of the Hilbert series (graded by the R-charge) of modules in an exact sequence must vanish, we can express the charge spectrum of $\cH_N$ in terms of an alternating sum of the charge spectra of $\cV_k$ over $k=0,1,2, \cdots$:
\begin{equation}
\Tr_{\cH_N} x^R = \sum_{k=0}^\infty (-1)^k \Tr_{\cV_k} x^R.
\end{equation}
This yields the half-BPS giant graviton expansion \eqref{eq:halfBPS gge bulk}. The closed string spectrum $Z_\infty = \prod_{n=1}^\infty \frac{1}{1-x^n}$ will factor out, because $\cV_k$ are free $R$-modules and $Z_\infty$ is the Hilbert series of $R$. This derivation shows that the resolution \eqref{eq:halfBPS resolution} ``categorifies'' the giant graviton expansion in the half-BPS sector.

What is the physical interpretation of the differential $\hQ$? It maps a half-BPS state of $k$ coincident D3 giants to linear combinations of states of $k-1$ coincident D3 giants with coefficients in the traces (i.e. closed strings). Thus a natural interpretation for $\hQ$ is that it is an instanton that interpolates between a pair of open string vacua with a ghost number difference of $1$. An example with $k=1$ would be a Euclidean D3 instanton on $S^3 \subset S^5$ described by the trajectory $\cos \theta(\tau) = e^{\tau - \tau_0}$ for $-\infty < \tau \leq \tau_0$, where $0 \leq \theta \leq \frac{\pi}{2}$ is a polar angle on $S^5$ that parametrizes the size of the wrapped $S^3$. It would be important to study the radiation due to such processes.

Our analysis here also clarifies why it is appropriate to describe half-BPS giant gravitons in terms of fermionic droplets, as is often done in the literature \cite{Lin:2004nb,Berenstein:2004kk,Dutta:2007ws}. We can collect the heavy ghosts into a single ``field''
\begin{equation}
\chi (z) = \sum_{n \in \mathbb{Z}} \frac{\chi_{n}}{z^{n+N+1}}
\end{equation}
on the $z$-plane of ghost number $1$. This is an anti-commuting field in the local transverse $z$-plane of $S^3$ in $S^5$ whose excitations are the states of D3 giants. It would be interesting to study the effective dynamics of giant gravitons along these lines.

The lessons from the half-BPS sector are the following: (1) we can identify the bulk space of states with the complex of auxiliary ghosts that compensate for finite $N$ trace relations, and (2) this complex furnishes a BRST-like resolution of the finite $N$ Hilbert space $\cH_N$ of the boundary gauge theory. These lessons are quite general when taken as statements about string theories dual to weakly-coupled gauge theories. Namely, they do not depend explicitly on supersymmetry or the presence of a weakly-curved gravity regime. That the bulk Hilbert space is a resolution of the gauge theory counterpart only requires the latter to be a theory of gauged $N \times N$ matrices.

\subsection{Mathematical background} \label{subsec:mathematicalbackground}

In this section, we develop the mathematical background required to generalize the lessons from the half-BPS sector. Some useful references are \cite{Henneaux:1992ig,Eisenbud:1995,Cox:2005,Felder:1988zp}. We present the background in the context that is relevant for this work.

\subsubsection{Finite $N$ modules and Koszul-Tate resolutions} \label{subsubsec:koszultate}

Consider a free four-dimensional $U(N)$ gauge theory on $\mathbb{R} \times S^3$ with fields only in the adjoint representation of the gauge group. We choose this manifold to avoid having topological sectors, but similar considerations apply to other manifolds as long as one works within a superselection sector for local operators.

In the limit $N \to \infty$, the space of states of the gauge theory is the ``vacuum'' free module $\cH_\infty = R |0\rangle$, given by the ring of gauge-invariant polynomials of fields and their derivatives
\begin{equation}
R = \mathbb{C}[\Tr X, \Tr X Y, \Tr \psi F_{\cdots}, \Tr \partial_{\cdots} \partial_{\cdots} X, \cdots].
\end{equation}
acting on the vacuum $|0 \rangle$. As an $R$-module, $\cH_\infty$ consists only of the identity generator acting on $|0 \rangle$ with coefficients in $R$. $\cH_\infty$ is graded with respect to the quantum numbers of the single trace generators of $R$. Following lessons from the half-BPS sector, we work over the ``large $N$'' ring $R$ generated by all single traces before taking into account the trace relations. In particular, $R$ is independent of $N$.

To find the space of states of a free gauge theory at finite $N$, one takes the quotient
\begin{equation}
\cH_N = \cH_\infty/\cI_N
\end{equation}
of $\cH_\infty$ by the submodule $\cI_N = I_N |0\rangle$, where the ideal over $R$
\begin{equation}
I_N = \langle \mathrm{trace \ relations \ at } \ N \rangle_R
\end{equation}
is generated by the set of trace relations that are present at a given value of $N$. For a single matrix $X$, these relations are given by the set $P_{N+1}, P_{N+2}, \cdots$ as provided explicitly in Section \ref{subsec:ghostsfortracerelations}. We refer to the quotient module $\cH_N$ over $R$ as the finite $N$ module. What makes the study of $\cH_N$ difficult is that $\cH_N$ is in general not freely generated, i.e. it is not a Fock space built from single trace generators. This means that in general we can only describe $\cH_N$ at best in terms of a set of generators that have relations among themselves.

The half-BPS sector suggests that, in order to find the ``bulk'' space of states for the string dual on a background labelled by $k$ wrapped D-branes, we should replace the quotient module $\cH_N = \cH_\infty / \cI_N$ by an infinite sequence of free $R$-modules $\cV_k^N$ that resolves $\cH_N$. The resulting complex is a free resolution of $\cH_N$, also known as the Koszul-Tate resolution.\footnote{We emphasize that this Koszul-Tate resolution, which implements trace relations as constraints on the large $N$ space of states, is distinct from the complex appearing in the BV-BRST quantization of gauge theories, where the resolution is used to implement the reduction to the on-shell surface. Therefore, the ``heavy'' ghosts that appear in this context are not the anti-ghosts of the BV-BRST formalism, although the mathematical tools are equivalent.} It is important that the resulting complex is a resolution of $\cH_N$, i.e. that the truncated complex as in \eqref{eq:halfBPS truncated complex} has homology only at ghost number $0$. Since the extra ghost degree is not present in $\cH_N$ of the boundary gauge theory, the presence of a non-vanishing homology at $k \neq 0$ would be at odds with holography.

A \textit{Koszul-Tate resolution} $\cK$ of $\cH_N$ is an exact sequence of free $R$-modules $\cV_k^{N}$ equipped with a nilpotent linear differential $\hQ$ 
\begin{equation}
\cdots \ \rightarrow \ \cV_3^{N} \ \xrightarrow{\hQ} \ \cV_2^{N} \ \xrightarrow{\hQ} \ \cV_1^{N} \ \xrightarrow{\hQ} \ \cH_\infty \rightarrow \cH_N \rightarrow 0
\end{equation}
where $\hQ \cV_1^{N} = \cI_N$, i.e. the image $\hQ \cV_1^{N}$ is the submodule generated by the trace relations at a given value of $N$. In addition to the grading by the quantum numbers induced from the generators of the ring $R$, the complex itself has a ``heavy'' ghost degree $\hgh$ under which
\begin{equation}
\hgh(\cV_k^N) = k, \qquad \hgh(\hQ) = -1.
\end{equation}
The generators of $\cV_k^{N}$ have quantum numbers that depend on $N$, because the set of trace relations depends on the value of $N$. The elements of $\cV_k^{N}$ are termed heavy because the energies/charges of their generators start at order $\sim k N$.

Intuitively, the resolution $\cK$ ``resolves'' the highly-constrained finite $N$ module $\cH_N$ by replacing it with an infinite sequence of Fock spaces $\cV_k^N$ (starting at $\cV_0 \equiv \cH_\infty$) that better and better approximate $\cH_N$. Since $\cH_\infty$ maps to $\cH_N$, the zeroth approximation to the finite $N$ module $\cH_N$ is the large $N$ module $\cH_\infty$. The subsequent free $R$-modules $\cV_k^N$ can be understood as the space of relations among the generators of $\cV_{k-1}^N$. For example, $\cV_1^N$ is the space of relations among the large $N$ Fock space generators\footnote{``Generators'' here refer to the generators of the ring $R$, rather than the identity generator of $\cH_\infty$ as an $R$-module. But when referring to the generators of $\cV_k^N$ for $k \geq 1$, we mean the generators of the $R$-module.} of $\cH_\infty$, i.e. the generators of $\cV_1$ map to trace relations under $\hQ$. Then $\cV_2^N$ is the space of relations among trace relations, and so on. Therefore, $\cV_k^N$ is the space of $k$-th order relations among the single-trace generators of the large $N$ ring $R$.

The resolution $\cK$ of $\cH_N$ consisting of free $R$-modules $\cV_k^N$ can always be constructed. For a given set of (commuting) quantum numbers $\{\cC_i\}$, the resolution $\cK$ truncates at finite length because there exists a value of $k$ beyond which all generators of $\cV_{k'}^N$ with $k'>k$ have quantum numbers that exceed $\{\cC_i\}$. It is important that $\cK$ has a notion of uniqueness despite having vanishing homology. We discuss its uniqueness properties in Section \ref{subsubsec:instantons}.

In physical terms, the Koszul-Tate resolution $\cK$ can be understood as the procedure of introducing heavy ghosts in a $U(\infty)$ theory that compensate for null states that would have arisen due to trace relations at a given value of $N$. There may be non-trivial relations between trace relations that require higher ghosts to compensate for them. Doing so in iteration reduces the space of states of a $U(\infty)$ gauge theory to that of a $U(N)$ gauge theory.

Explicitly, let $P_A \in R$ be a trace relation at $N$, where the label $A$ ranges over the (infinite) set of trace relations. For every $P_A$, we introduce a heavy ghost $\chi_A$ of $\hgh(\chi_A) = 1$ with opposite grassmann parity to that of $P_A$. The nilpotent differential $\hcQ = [ \, \cdot \,, \hQ]_\pm$ on operators acts as
\begin{equation}
\hcQ \chi_{A} = P_A,
\end{equation}
which makes $P_A$ $\hcQ$-exact.

The set of trace relations $P_A$ may not be linearly-independent over the ring $R$, i.e. $P_A$ may satisfy
\begin{equation} \label{eq:relation between trace relations}
r_1 P_{A_1} + r_2 P_{A_2} + \cdots + r_s P_{A_s} = 0
\end{equation}
where $r_i$ are non-zero elements of $R$. \eqref{eq:relation between trace relations} is a relation between trace relations that needs to compensated via ghosts. Actually, there always exist relations between trace relations of the form
\begin{equation}
P_{A_2} P_{A_1} -P_{A_1} P_{A_2} = 0
\end{equation}
where we set $r_1 = P_{A_2}, \, r_2 = -P_{A_1}$ and assumed $P_A$ are grassmann even. This is a ``trivial'' relation because it simply states that the trace relations $P_{A_1}, P_{A_2}$ commute. Trivial relations that result from the (anti-)commutation properties of trace relations are automatically accounted for by taking $\hcQ$ to act on products of ghosts via the graded Leibniz rule:
\begin{align} \label{eq:trivial relations q-exact}
    \hcQ ( \chi_{A_1} \chi_{A_2} ) &= P_{A_1} \chi_{A_2} + (-1)^{\varepsilon_1} P_{A_2} \chi_{A_1} \nonumber \\
    \hcQ (\chi_{A_1} \chi_{A_2} \chi_{A_3}) &= P_{A_1} \chi_{A_2} \chi_{A_3} + (-1)^{\varepsilon_1} P_{A_2} \chi_{A_1} \chi_{A_3} + (-1)^{\varepsilon_1 + \varepsilon_2} P_{A_3} \chi_{A_1} \chi_{A_2} 
\end{align}
and so on, where $\varepsilon_i=0$ if $\chi_{A_i}$ is grassmann even and $\varepsilon_i=1$ if $\chi_{A_i}$ is grassmann odd. Thus, the (anti-)commutation properties are implemented by the statement that the expression on the RHS of \eqref{eq:trivial relations q-exact} are $\hcQ$-exact. If there are only trivial relations, it is sufficient to consider all multi-ghost generators
\begin{equation}
\chi_{A_1} \chi_{A_2} \cdots \chi_{A_k} |0\rangle \ \in \ \cV_k^N
\end{equation}
with coefficients in $R$, and the procedure of finding the Koszul-Tate resolution $\cK$ terminates.

However, there may exist non-trivial relations \eqref{eq:relation between trace relations} between trace relations $P_A$ that do not just amount to the (anti-)commutation properties of $P_A$. In this case, we must introduce ghosts-for-ghosts $\chi^{(2)}$ of $\hgh(\chi^{(2)}) = 2$ satisfying
\begin{equation}
\hcQ ( \chi^{(2)}_{B} ) = r_1 \chi_{A_1}^{(1)} + r_2 \chi_{A_2}^{(1)} + \cdots + r_s \chi_{A_s}^{(1)},
\end{equation}
where $\chi^{(1)}_{A_i} \equiv \chi_{A_i}$. This implements non-trivial relations between trace relations, and the subscript $B$ here labels the set of such relations. If there are no further non-trivial relations (again with coefficients in $R$) between the various relations \eqref{eq:relation between trace relations}, the procedure of finding the Koszul-Tate resolution $\cK$ terminates. If higher relations do exist, the process iterates until one has accounted for all non-trivial higher relations by making them $\hcQ$-exact with respect to $\chi^{(h)}$ of $\hgh(\chi^{(h)}) = h$.\footnote{For example, if there exists a relation between relations between trace relations, then the compensating ghost $\chi^{(3)}$ maps to a linear combination of $\chi_{B}^{(2)}$ and $\chi_{A_1}^{(1)}\chi_{A_2}^{(1)}$ with coefficients in $R$.} Given a set of quantum numbers $\{ \cC_i \}$, there are only finitely many generators and relations with charges less than or equal to $\{ \cC_i \}$.

In general, the differential $\hcQ$ on operators acts as
\begin{equation}
\hcQ \left( \chi^{(h_1)}_{A_1} \chi^{(h_2)}_{A_2} \cdots \chi^{(h_n)}_{A_n} \right) = \sum_{i=1}^n (-1)^{\varepsilon_1+ \varepsilon_2 + \cdots + \varepsilon_{i-1}} \chi^{(h_1)}_{A_1} \chi^{(h_2)}_{A_2} \cdots \hcQ(\chi^{(h_i)}_{A_i}) \cdots \chi^{(h_n)}_{A_n}
\end{equation}
where $\varepsilon_i=0$ if $\chi_{A_i}^{(h_i)}$ is grassmann even and $\varepsilon_i=1$ if $\chi_{A_i}^{(h_i)}$ is grassmann odd. The Koszul-Tate resolution is a complex built from multi-particling all such ghosts $\chi_{A_i}^{(h_i)}$ with coefficients in the large $N$ ring $R$. States of total ghost number $k$
\begin{equation}
\chi^{(h_1)}_{A_1} \chi^{(h_2)}_{A_2} \cdots \chi^{(h_n)}_{A_n} |0\rangle \ \in \ \cV_k^N
\end{equation}
generate the free $R$-module $\cV_k^N$ where $k = \sum_{i=1}^n h_i$.

Once a resolution $\cK$ is constructed, we can consider the truncated complex $\cV^N$:
\begin{equation}
\cdots \ \rightarrow \ \cV_3^{N} \ \xrightarrow{\hQ} \ \cV_2^{N} \ \xrightarrow{\hQ} \ \cV_1^{N} \ \xrightarrow{\hQ} \ \cH_\infty \rightarrow 0
\end{equation}
whose homology is concentrated at heavy ghost number $0$:
\begin{equation}
\left( \frac{\mathrm{Ker}\, \hQ}{\mathrm{Im}\, \hQ} \right)_{\hgh=0} = \, \cH_N.
\end{equation}
So the complex $\cV^N$ of free $R$-modules encodes the data of the finite $N$ module $\cH_N$ via its homology. We also refer to the complex $\cV^N$ as a resolution of $\cH_N$.

We find evidence in Section \ref{sec:basicexamples} that higher ghosts are required even in rather simple settings, such as the free $\frac{1}{4}$-BPS sector and the $\frac{1}{4}$-BPS chiral ring of $\Nfour$ SYM. It would be important to understand the criteria for when they arise and, in situations where they exist, the holographic description of higher ghosts.

\subsubsection{Thermal partition function and correlators} \label{subsubsec:lefschetztrace}

Given a Koszul-Tate resolution $\cK$, the Lefschetz trace formula \cite{Henneaux:1992ig} tells us how to relate physical quantities computed over $\cH_N$ and over its resolution $\cV^N$:
\begin{equation} \label{eq:lefschetz}
\Tr_{\cH_N} \cO = \sum_{k=0}^\infty (-1)^k \Tr_{\cV_k^N} \cO
\end{equation}
where $\cO$ is an operator of heavy ghost number $0$ in the $U(N)$ gauge theory that we also take to act on $\cV_k^N$. To compute the trace, we need the concept of a dual module $\left(\cV_k^N\right)^*$, whose elements are related to those of $\cV_k^N$ by the hermitian adjoint and carry opposite heavy ghost number $\hgh$ and quantum numbers $\{ \cC_i \}$ to those of $\cV_k^N$. We work in an orthonormal basis for $\left(\cV_k^N\right)^*$ and $\cV_k^N$ of definite energy and charge eigenvalues.

The Lefschetz trace formula will provide the holographic map in our proposal of Section \ref{subsec:generalbulkspace}, in the regime of $\lambda = 0$ and finite $N$. Bulk observables are computed by taking the alternating sum of the expectation values in an ensemble of states built on each open string vacuum. For instance, the free thermal partition function of a $U(N)$ gauge theory on $S^1 \times S^3$ is equal to the alternating sum of the free thermal partition functions over $\cV_k^N$ of the resolution $\cV^N$:
\begin{equation} \label{eq:lefschetzthermalpartition}
\Tr_{\cH_N} e^{-\beta H} = \sum_{k=0}^\infty (-1)^k \Tr_{\cV_k^N} e^{-\beta H}.
\end{equation}
The thermal correlation functions are related as
\begin{align} \label{eq:lefschetzcorrelator}
    \Tr_{\cH_N} &\left[ e^{-\beta H} \, \cO_1 (\tau_1, x_1) \cO_2 (\tau_2, x_2) \cdots \cO_n(\tau_n,x_n) \right] \nonumber \\
    &= \sum_{k=0}^\infty (-1)^k \Tr_{\cV_k^N} \left[ e^{-\beta H} \, \cO_1 (\tau_1, x_1) \cO_2 (\tau_2, x_2) \cdots \cO_n(\tau_n,x_n) \right]
\end{align}
where $0 \leq \tau_n < \cdots < \tau_2 < \tau_1 < \beta$. The hamiltonian on the right-hand side is defined in the $U(\infty)$ theory supplemented with ghosts for trace relations at a value of $N$.

For superconformal gauge theories with non-anomalous R-symmetry, we can write in addition the relation between the superconformal indices
\begin{equation} \label{eq:lefschetzindices}
\Tr_{\cH_N} \left[ (-1)^F e^{-\beta \Delta} q_1^{\cC_1} q_2^{\cC_2} \cdots \right] = \sum_{k=0}^\infty (-1)^k \Tr_{\cV_k^N} \left[ (-1)^F e^{-\beta \Delta} q_1^{\cC_1} q_2^{\cC_2} \cdots \right]
\end{equation}
where $\Delta = \frac{1}{2} \{ \cQ, \cQ^\dag \} = 0$ is the BPS condition and $\cC_i$ are conserved charges dual to fugacities $q_i$ that commute with the supercharge $\cQ$. In Section \ref{subsubsec:largeN}, we describe the conditions under which the dependence of the right-hand side on $N$ simplifies and \eqref{eq:lefschetzindices} may be expressed as
\begin{equation} \label{eq:gge lefschetz}
Z_N(x, y_i) = Z_\infty(x, y_i) \sum_{k=0}^\infty x^{k N} \hat{Z}_k(x,y_i),
\end{equation}
which is the giant graviton expansion proposed in \cite{Gaiotto:2021xce}. Various minus signs from both $(-1)^k$ and $(-1)^F$ are hidden in the definition of $\hat{Z}_k(x,y_i)$. If the bulk has a supergravity regime, the indices on the right of \eqref{eq:gge lefschetz} can also be computed directly in the bulk by counting the fluctuation modes of probe D-branes \cite{Arai:2019xmp,Imamura:2021ytr}.

\subsubsection{(Non-)uniqueness and instantons} \label{subsubsec:instantons}

Here we comment on the uniqueness properties of a Koszul-Tate resolution $\cK$ of $\cH_N$. First, we discuss why a Koszul-Tate resolution $\cK$ of $\cH_N$ does not necessarily produce a unique set of free $R$-modules $\cV_k^{N}$. Then we explain that there nonetheless exists a notion of uniqueness for $\cK$, which we use to make physical statements about the bulk spectrum in Section \ref{subsubsec:lowerbound}.

It is not hard to understand why there exist infinitely many resolutions of $\cH_N$ at any finite value of $N$. Given a resolution $\cK$, one may always introduce an additional pair of spurious heavy ghosts $\xi^{(k)}$ and $\xi^{(k-1)}$ on which the differential $\hcQ$ act as
\begin{align} \label{eq:spurious ghosts}
    &\hcQ \xi^{(k)} = \xi^{(k-1)} \nonumber \\
    &\hcQ \xi^{(k-1)} = 0.
\end{align}
Contributions from such pairs cancel in the Lefschetz trace formula \eqref{eq:lefschetz} due to the relative minus sign $(-1)^k$. So an arbitrary number of them can be introduced without affecting physical quantities computed by \eqref{eq:lefschetz}. However, doing so still changes the energy/charge spectra associated with individual pairs of free $R$-modules $\cV_k^{N}$ and $\cV_{k-1}^{N}$.

This situation is analogous to the effect, on the instanton complex in supersymmetric quantum mechanics, of deforming the Morse function $h$ on the target manifold such that the number of critical points of $h$ changes \cite{Witten:1982im}. Under generic deformations, the number of non-degenerate critical points changes by two, and the created/annihilated pair of critical points have Morse indices that differ by $1$. While the number of critical points of $h$ with a given Morse index $k$ may change by an arbitrary positive amount, the Euler characteristic remains invariant. The difference between the instanton complex and our resolution $\cV^N$, however, is that the instanton complex may have non-trivial cohomology at any form degree while the homology of $\cV^N$ is concentrated only at heavy ghost number $0$.

Following the analogy with the instanton complex, our interpretation of the situation as follows: At any positive integer $N$, there is an inherent ambiguity in the spectrum associated to a string background labelled by a collection of $k$ wrapped D-branes, due to the unphysical freedom to introduce spurious pairs of ghosts \eqref{eq:spurious ghosts} between D-brane sectors $\cV_k^{N}$ and $\cV_{k-1}^{N}$ whose ghost numbers differ by $1$. Thus, while the number of states with a given ghost degree $k$ and quantum numbers $\{ \cC_i \}$ may change by an arbitrary positive amount, any physical quantity computed by the Lefschetz trace formula remains invariant. 

It is still interesting to ask whether there are physical ``invariants'' of $\cK$ that are non-trivial at each ghost number $k$. This is because in the bulk we can at best compute only on backgrounds with a fixed set of D-branes. One would like to be able to compare bulk quantities on a background with $k$ wrapped D-branes with the trace over $\cV_k^{N}$ found via the Koszul-Tate resolution $\cK$ of $\cH_N$. It turns out that there is indeed a sense in which the modules $\cV_k^N$ of $\cK$ are uniquely determined:

A \textit{minimal} Koszul-Tate resolution $\hcK$ (also known as a minimal graded free resolution) is that for which the number of generators of each free $R$-module $\cV_k^N$ in $\cK$ is minimal \cite{Eisenbud:1995}, i.e. a resolution $\hcK$ for which there are no spurious ghosts. Though the modules $\cV_k^N$ possess an infinite number of generators, there are only a finite number of generators with a given set of quantum numbers $\{ \cC_i \}$. Therefore, the minimality of $\hcK$ is well-defined for any commuting set $\{ \cC_i \}$. A minimal Koszul-Tate resolution $\hcK$ of $\cH_N$ is unique up to isomorphisms, and any other non-minimal resolution of $\cH_N$ may be constructed by taking the direct sum of $\hcK$ with a trivial complex consisting of spurious ghosts.

\subsection{Bulk space of states at $\lambda = 0$} \label{subsec:generalbulkspace}

In Section \ref{subsec:halfbpsbulk}, we provided strong evidence for the following statement in the half-BPS sectors of $U(N)$ $\Nfour$ SYM and IIB string theory on $\AdStimesS$:
\begin{equation} \label{eq:halfBPS equality}
\cV_k = \cH_\infty^{\mathrm{closed}} \otimes \cH_k^{\mathrm{open}} \quad (\textrm{half-BPS}).
\end{equation}
In words, (1) the free $R$-module $\cV_k$ consisting of states of ghost degree $k$ in a resolution \eqref{eq:halfBPS resolution} of the half-BPS Hilbert space $\cH_N$ at $\lambda=0$ is holographically dual to (2) the half-BPS open-closed Hilbert space $\cH_\infty^{\mathrm{closed}} \otimes \cH_k^{\mathrm{open}}$ of IIB strings on $\AdStimesS$ with $k$ coincident D3 giant graviton branes, in the $\alpha' \to \infty$ limit. It is possible to make a statement such as \eqref{eq:halfBPS equality}, where there exists a sharp semiclassical bulk description, because of special properties of the half-BPS resolution that we clarify in Section \ref{subsubsec:largeN}.

In more general circumstances, $\cV_k^{N}$ of the Koszul-Tate resolution $\cK$ generalizes the bulk interpretation associated with $\cH_\infty^{\mathrm{closed}} \otimes \cH_k^{\mathrm{open}}$ in a few important ways. In the limit where $N$ is large and all but one $U(1)$ charge is held fixed, the module $\cV_k^N$ asymptotes to (in a sense that will be made precise) the space of string states on $k$ coincident D-branes fixed under the $U(1)$ bulk isometry. This regime includes non-BPS excitations of the half-BPS sector. In other regimes, such as when two or more charges are treated on an equal footing, $\cV_k^N$ ceases to have a ``good'' large $N$ limit: the number of elements of $\cV_k^N$ of a given total charge or energy (minus the total background charge $k N$) diverges as $N \to \infty$. The module $\cV_k^N$ nonetheless remains well-defined at any finite value of $N$ where no semiclassical bulk description may exist, modulo unphysical ambiguities described in Section \ref{subsubsec:instantons}. It is in this sense that $\cV_k^{N}$ generalizes the notion of the ``large curvature limit of the space of string states on a background with $k$ wrapped D-branes''.

\paragraph{The proposal}

Consider a four-dimensional $U(N)$ gauge theory on $\mathbb{R} \times S^3$ possessing only fields in the adjoint representation of the gauge group, at zero 't Hooft coupling $\lambda = 0$. Let $R$ be the ring of polynomials of traces of fields and their derivatives (before accounting for trace relations). The space of states $\cH_N$ of the gauge theory is the quotient of $\cH_\infty = R |0\rangle$ by the submodule generated by the set of all trace relations at a given value of $N$.

\textit{Our proposal is that the complex $\cV^N$:}
\begin{equation}
\cdots \ \rightarrow \ \cV_3^{N} \ \xrightarrow{\hQ} \ \cV_2^{N} \ \xrightarrow{\hQ} \ \cV_1^{N} \ \xrightarrow{\hQ} \ \cH_\infty \rightarrow 0,
\end{equation}
\textit{which furnishes a Koszul-Tate resolution $\cK$ of $\cH_N$ over the ring $R$, should be identified as the space of states of the dual string theory in the $\alpha' \to \infty$ limit.} The homology of $\cV^N$ is concentrated at heavy ghost number $0$ and coincides with $\cH_N = \cH_\infty / \hQ \cV_1^N$. The free $R$-module $\cV_k^N$ is, in a sense that is elaborated, the $\alpha' \to \infty$ limit of the space of string states on a background with $k$ wrapped D-branes. The observables of the $U(N)$ gauge theory are related to those of its string dual via the Lefschetz trace formula
\begin{equation} \label{eq:lefschetz proposal}
\Tr_{\cH_N} \cO = \sum_{k=0}^\infty (-1)^k \Tr_{\cV_k^N} \cO
\end{equation}
which provides the holographic map in the regime of $\lambda = 0$ and finite $N$. Bulk observables are computed by taking the alternating sum of the expectation values in an ensemble of states built on each open string vacuum.

There exist instanton-like ambiguities for the modules $\cV_k^N$ at finite $N$ that arise from the freedom to introduce spurious ghosts in the resolution without affecting physical quantities computed via \eqref{eq:lefschetz proposal}. We can nonetheless place lower bounds on the spectrum of string states on D-brane backgrounds at any value of $N$, using the uniqueness of minimal resolutions. In the limit where $N$ is large and all but one $U(1)$ charge is held fixed, we can determine exactly the string spectrum on certain D-branes in the large curvature limit from the counting of trace relations in a $U(N)$ gauge theory.

It is clear that various assumptions for the gauge theory may be relaxed to produce interesting outcomes. We comment on these possibilities in the discussion.

\subsubsection{Semiclassics and large $N$} \label{subsubsec:largeN}

Let us discuss the circumstances under which the module $\cV_k^N$ acquires a sharp bulk description as the space of states of $k$ coincident giant graviton-like branes. In this limit, the string spectrum on D-brane backgrounds in the large curvature limit can be determined from the counting of trace relations in the $U(N)$ gauge theory.

We begin by stating some reasonable properties for this regime: For one, $N$ must be large. Moreover, as we saw in Section \ref{subsec:localization}, the D-branes that contribute states in the large curvature limit $L \to 0$ wrap the fixed point of a $U(1)$ bulk isometry. They also possess large charges $\sim k N$ under this $U(1)$ due to the induced potential. Therefore, $\cV_k^N$ is well-described as the space of string states on $k$ coincident D-branes at a $U(1)$ fixed-point when $N$ is large and when all but one $U(1)$ charge are held fixed.

This limit corresponds to considering trace relations that contain an arbitrary number of a single adjoint field $X$ and fixed numbers of other adjoint ``letters'' $W_1, W_2, \cdots$. For instance assuming all letters are bosonic, the set of trace relations for $N=2$ and containing two letters $W_1, W_2$ starts at
\begin{align}
    &\Tr X W_1 W_2 + \Tr X W_2 W_1  - \Tr X \, \Tr W_1 W_2 \nonumber  \\
    &- \Tr W_1\, \Tr X W_2 - \Tr W_2\, \Tr X W_1  + \Tr X\, \Tr W_1\, \Tr W_2 = 0.
\end{align}
There is $1$ relation with $W_1, W_2$ and one $X$. Next, we have
\begin{align}
    &2\, \Tr X^2 W_1 W_2 - 2\, \Tr X\, \Tr X W_1 W_2 - \Tr X^2\, \Tr W_1 W_2 + (\Tr X)^2\, \Tr W_1 W_2 = 0 \nonumber \\
    &2\, \Tr X^2 W_2 W_1 + 2\, \Tr X\, \Tr X W_1 W_2 - \Tr X^2\, \Tr W_1 W_2 - (\Tr X)^2\, \Tr W_1 W_2 \nonumber \\
    &- 2\, \Tr X\, \Tr W_1\, \Tr X W_2 - 2\, \Tr X\, \Tr W_2\, \Tr X W_1 + 2\, (\Tr X)^2\, \Tr W_1\, \Tr W_2 = 0 \nonumber \\
    &2\, \Tr X W_1 X W_2 + \Tr X^2\, \Tr W_1 W_2 - (\Tr X)^2\, \Tr W_1 W_2 \nonumber \\
    &- 2\, \Tr X W_1\, \Tr X W_2- \Tr X^2\, \Tr W_1\, \Tr W_2 + (\Tr X)^2\, \Tr W_1\, \Tr W_2 = 0.
\end{align}
There are $3$ independent relations with $W_1, W_2$ and two $X$s, and so on with higher insertions of $X$s. 

What do we mean by the large $N$ limit of the space of trace relations? Let us define
\begin{equation}
Z_{\cV_k^N} = \Tr_{\cV_k^N} \left[ x^{\cC_X} w_1^{\cC_1} w_2^{\cC_2} \cdots \right]
\end{equation}
where $\cC_{i}$ are charges of the letters $W_{i}$.\footnote{The letters $W_i$ may also carry $\cC_X$ charges. We can specialize $w_i$ and $\cC_i$ later as long as there are a finite number of $W_i$ with fixed charges.} Since $\cV_k^N$ are free $R$-modules, we can divide $Z_{\cV_k^N}$ by the large $N$ partition function
\begin{equation}
Z_\infty = \Tr_{\cH_\infty} \left[ x^{\cC_X} w_1^{\cC_1} w_2^{\cC_2} \cdots \right]
\end{equation}
to find the quantum numbers of the generators of $\cV_k^N$. Since the module $\cV_{1}^N$ of a minimal Koszul-Tate resolution is isomorphic to the space of trace relations with coefficients in $R$, we are interested in the large $N$ behavior of the generators of $\cV_{1}^N$ that map to trace relations containing the letters $W_1, W_2$. As $N$ is increased, the number of such generators of the free $R$-module $\cV_{1}^N$ stabilizes up to an overall shift in $\cC_X$ by $N$ units:
\begin{equation} \label{eq:largeN G1}
\frac{Z_{\cV_{1}^{N \to \infty}}}{Z_\infty} \bigg\rvert_{w_1 w_2} = \ x^{N-1} + 3 x^{N} + 5 x^{N+1} + 7 x^{N+2} + 9 x^{N+3} + 11 x^{N+4} + 13 x^{N+5} + \cdots
\end{equation}
where $\rvert_{w_1^{\cC_1} w_2^{\cC_2} \cdots}$ denotes that we take the coefficient of $w_1^{\cC_1} w_2^{\cC_2} \cdots$. The first two coefficients of the power series in $x$ agree with the counting of independent trace relations at $N=2$ in the preceding paragraph, but the higher coefficients do not agree:
\begin{equation}
\frac{Z_{\cV_{1}^{N=2}}}{Z_\infty} \bigg\rvert_{w_1 w_2} = \ x + 3 x^{2} + 4 x^{3} + 5 x^{4} + 6 x^{5} + 7 x^{6} + 8 x^{7} + \cdots.
\end{equation}
Our claim is that these series coefficients converge to those of \eqref{eq:largeN G1} as one considers modules $\cV_{1}^N$ in a sequence of minimal Koszul-Tate resolutions of increasing $N$. As another example, we write the large $N$ limit of the number of independent trace relations that contain letters $W_1, W_2, W_3$:
\begin{equation}
\frac{Z_{\cV_{1}^{N \to \infty}}}{Z_\infty} \bigg\rvert_{w_1 w_2 w_3} = \ x^{N-2} + 7 x^{N-1} + 19 x^{N} + 37 x^{N+1} + 61 x^{N+2} + 91 x^{N+3} + \cdots.
\end{equation}

The set of trace relations possesses degeneracies in the $U(1)$ charge that stabilizes at large $N$, with other charges $\cC_i$ fixed. As such, we observe that
\begin{equation} \label{eq:G1 definition}
\lim_{N \to \infty} \cV_{1}^N \Big\rvert_{\cC_i} \ = \ \cG_{1} \Big\rvert_{\cC_i},
\end{equation}
namely that the space $\cV_{1}^N$ of (ghosts for) trace relations converges in the large $N$ limit to another free $R$-module $\cG_{1}$ that depends trivially on $N$. We conjecture that generators of $\cG_{1}$ are holographically dual to states of a single D-brane on the $U(1)$ fixed locus in the $\alpha' \to \infty$ limit of the string dual.

This observation is actually just the first step at $k=1$ of the giant graviton expansion \cite{Gaiotto:2021xce}
\begin{equation} \label{eq:gge largeN}
Z_N(x, w_i) = Z_\infty(x, w_i) \sum_{k=0}^\infty x^{k N} \hat{Z}_k(x,w_i),
\end{equation}
for superconformal indices when all but one $U(1)$ charge are held fixed. But now we have an understanding of the $k=1$ term of \eqref{eq:gge largeN} at the level of trace relations and resolution thereof:
\begin{equation} \label{eq:G1 identification}
x^{N} Z_\infty(x, w_i) \hat{Z}_{1}(x,w_i) = -\Tr_{\cG_{1}} \left[ (-1)^F x^{\cC_X} w_1^{\cC_1} w_2^{\cC_2} \cdots \right].
\end{equation}
The minus sign on the right side compensates for an overall minus sign that arises from the definition of $\hat{Z}_{1}$ due to an analytic continuation of the fugacity $x$. The identification \eqref{eq:G1 identification} holds when the letters $W_i$ preserve supersymmetries used the define the index. Note however that our definition \eqref{eq:G1 definition} of $\cG_1$ based on trace relations does not require that $W_i$ are BPS letters.

An important property of the formula \eqref{eq:gge largeN} is that the right hand side depends on $N$ only through the prefactor $x^{k N}$, i.e. $\hat{Z}_k(x,w_i)$ are independent of $N$. This suggests that, if we add the number of generators of $\cG_1$ to
\begin{equation}
\frac{1}{Z_\infty} \Tr_{\cH_N} \left[ x^{\cC_X} w_1^{\cC_1} w_2^{\cC_2} \cdots \right] - 1,
\end{equation}
the remaining series starting at $O(x^{2N})$ also stabilizes in the large $N$ limit. We take the remaining series to count (up to overall sign) the number of generators of another free $R$-module $\cG_2$. For example, the generators of $\cG_2$ when the trace relations involve arbitrary numbers of $X$ and fixed numbers of letters $W_1, W_2$ and $W_1,W_2,W_3$ possess the series
\begin{align}
    \frac{Z_{\cG_{2}}}{Z_\infty} \bigg\rvert_{w_1 w_2} &= \ x^{2N-1} + 4 x^{2N} + 9 x^{2N+1} + 16 x^{2N+2} + 25 x^{2N+3} + 36 x^{2N+4} + \cdots \nonumber \\
    \frac{Z_{\cG_{2}}}{Z_\infty} \bigg\rvert_{w_1 w_2 w_3} &= \ x^{2N-3} + 6 x^{2N-2} + 22 x^{2N-1} + 56 x^{2N} + 114 x^{2N+1} + 202 x^{2N+2} + \cdots.
\end{align}
As the procedure may be continued iteratively, this suggests that there exists a non-minimal Koszul-Tate resolution $\cG$ of $\cH_N$
\begin{equation}
\cdots \ \rightarrow \ \cG_3 \ \xrightarrow{\hQ} \ \cG_2 \ \xrightarrow{\hQ} \ \cG_1 \ \xrightarrow{\hQ} \ \cH_\infty \rightarrow \cH_N \rightarrow 0
\end{equation}
at any value of $N$ (again understood within each $\cC_i$-eigenspace), where the free $R$-modules $\cG_k$ depend on $N$ only via an overall shift in the $\cC_X$ charges of its generators.\footnote{Note that in general $\cG_k \big\rvert_{\cC_i} \neq \lim_{N \to \infty} \cV_k^N \big\rvert_{\cC_i}$ for $k>1$.}

The non-minimal resolution $\cG$ generalizes the expansion \eqref{eq:gge largeN} from a statement about superconformal indices
\begin{equation}
\Tr_{\cH_N} \left[ (-1)^F x^{\cC_X} w_1^{\cC_1} w_2^{\cC_2} \cdots \right] = \sum_{k=0}^\infty (-1)^k \Tr_{\cG_k} \left[ (-1)^F x^{\cC_X} w_1^{\cC_1} w_2^{\cC_2} \cdots \right],
\end{equation}
to that about the charge spectrum of free gauge theories at finite $N$
\begin{equation}
\Tr_{\cH_N} \left[ x^{\cC_X} w_1^{\cC_1} w_2^{\cC_2} \cdots \right] = \sum_{k=0}^\infty (-1)^k \Tr_{\cG_k} \left[ x^{\cC_X} w_1^{\cC_1} w_2^{\cC_2} \cdots \right],
\end{equation}
because $\cG_k$ can be found only based on the analysis of trace relations. We conjecture that the generators of the free $R$-module $\cG_{k}$ are holographically dual to states of $k$ coincident D-branes on a fixed locus of the $U(1)$ isometry in the $\alpha' \to \infty$ limit of the dual string theory.

\subsubsection{A lower bound on D-brane states} \label{subsubsec:lowerbound}

At finite $N$ and generic charges, the bulk description of the module $\cV_k^N$ as the space of string states on a background with $k$ wrapped D-branes becomes less precise. A reflection of this fact on $\cV_k^N$ is that the number of states of $\cV_k^N$ of a given \textit{total} charge or energy (minus the total background charge $k N$) necessarily diverges as $N \to \infty$, in contrast to the situation in the previous section. So in general the space $\cV_k^N$ does not have a ``good'' large $N$ limit.

Nevertheless, the space $\cV_k^N$ in a Koszul-Tate resolution of $\cH_N$ remains well-defined at any finite value of $N$ and energies, because $\cV_k^N$ is defined through the set of trace relations and their relations in a boundary $U(N)$ gauge theory. It is therefore possible to put constraints on the ``quantum'' D-brane sectors based on our knowledge of $\cV_k^N$.

According to our proposal, the free $R$-module $\cV_k^N$ in $\cK$ gives the spectrum of string states on $k$ D-brane background. Due to the freedom to add spurious pairs of ghosts between $\cV_k^N$ and $\cV_{k-1}^N$, this spectrum is in principle subject only to the global constraint
\begin{equation} \label{eq:globalconstraint}
\Tr_{\cH_N} e^{-\beta H} = \sum_{k=0}^\infty (-1)^k \Tr_{\cV_k^N} e^{-\beta H}.
\end{equation}
We interpret this freedom as an indication that, at finite $N$, there may exist different computational schemes in the bulk for computing the spectrum on D-brane backgrounds, where the different schemes are required only to satisfy the global constraint \eqref{eq:globalconstraint}. We discussed an analogy of this situation to that of the instanton complex in Section \ref{subsubsec:instantons}.

Let us show that it is still possible to make useful statements about the spaces $\cV_k^N$ at each degree $k$. Recall from Section \ref{subsubsec:instantons} that a minimal Koszul-Tate resolution $\hcK$ of $\cH_N$ is unique, i.e. any non-minimal resolution $\cK$ of $\cH_N$ may be constructed by adjoining to $\hcK$ spurious pairs of ghosts that decouple from $\hcK$. Given the set of trace relations in a $U(N)$ gauge theory, we use the uniqueness of $\hcK$ to place lower bounds on the spectrum of string states on D-brane backgrounds in the string dual at $\alpha' \to \infty$ and finite $N$.

Let $\hcK$ and $\cK$, respectively, be minimal and non-minimal Koszul-Tate resolutions of $\cH_N$:
\begin{align}
    \hcK \ &: \ \cdots \ \rightarrow \ \widehat\cV_3^{N} \ \xrightarrow{\hQ} \ \widehat\cV_2^{N} \ \xrightarrow{\hQ} \ \widehat\cV_1^{N} \ \xrightarrow{\hQ} \ \cH_\infty \rightarrow \cH_N \rightarrow 0 \nonumber \\
    \cK \ &: \ \cdots \ \rightarrow \ \cV_3^{N} \ \xrightarrow{\hQ} \ \cV_2^{N} \ \xrightarrow{\hQ} \ \cV_1^{N} \ \xrightarrow{\hQ} \ \cH_\infty \rightarrow \cH_N \rightarrow 0,
\end{align}
where $\hQ$ are defined differently in $\hcK$ and $\cK$. The Lefschetz trace formula says that
\begin{equation}
\Tr_{\cH_N} e^{-\beta H} = \sum_{k=0}^\infty (-1)^k \Tr_{\widehat\cV_k^N} e^{-\beta H} = \sum_{k=0}^\infty (-1)^k \Tr_{\cV_k^N} e^{-\beta H},
\end{equation}
which may be further refined by global charges. If we define the Hilbert-Poincar\'{e} series
\begin{align}
    \widehat{\mathbf{P}}(\mu, q) &= \sum_{k=0}^\infty \mu^k \, \Tr_{\widehat\cV_k^N} e^{-\beta H} \nonumber \\
    \mathbf{P}(\mu, q) &= \sum_{k=0}^\infty \mu^k \, \Tr_{\cV_k^N} e^{-\beta H}
\end{align}
where $q = e^{-\beta}$, then by the uniqueness of the minimal resolution $\hcK$, there exists $\mathbf{Q}(\mu,q)$ with non-negative coefficients in the $\mu$- and $q$-series such that the following holds:
\begin{equation} \label{eq:lowerbound}
\mathbf{P}(\mu, q) - \widehat{\mathbf{P}}(\mu, q) = (1 + \mu) \mathbf{Q}(\mu, q).
\end{equation}
This is the statement that $\cK$ can be constructed from a minimal resolution $\hcK$ by introducing spurious \textit{pairs} of heavy ghosts $\xi^{(k)}$ and $\xi^{(k-1)}$ satisfying $\hcQ \xi^{(k)} = \xi^{(k-1)}$ and $\hcQ \xi^{(k-1)} = 0$. The ghosts $\xi^{(k)}$ and $\xi^{(k-1)}$ possess the same energy eigenvalue. \eqref{eq:lowerbound} places a lower bound at each energy eigenvalue on the number of generators of $\cV_k^N$, i.e. on the degeneracy of open strings on D-branes in the $\alpha' \to \infty$ limit.

Note that the lower bound \eqref{eq:lowerbound} applies when there is a state-counting interpretation for the bulk partition function computed in background with $k$ D-branes. For instance, one situation where the bound does not apply is in the giant graviton-like expansion considered in \cite{Murthy:2022ien}. It was found in \cite{Eniceicu:2023uvd} that brane partition functions in this expansion originate from a $U(m|m)$ super-matrix integral, which suggests that the partition function is associated to a more exotic brane system. It would be interesting to understand the expansion of \cite{Murthy:2022ien} in terms of Koszul-Tate resolutions.

Another situation where the bound does not apply is when analytic properties of the bulk partition functions are used to find the spectrum. For example, the giant graviton expansion for the $\frac{1}{4}$-BPS chiral ring of $U(N)$ $\Nfour$ SYM can be found by summing over residues of a grand canonical partition function:
\begin{align} \label{eq:quarterBPS gge}
    Z_N(x,y) &= Z_\infty \sum_{k,k' = 0}^\infty \frac{x^{k N} y^{k' N}}{\prod_{n=1}^{k} \prod_{m=0}^{\infty} (1 - x^{-n} y^m) \prod_{n=1}^{k'} \prod_{m=0}^{\infty} (1 - y^{-n} x^m) \prod_{n=1}^{k} \prod_{m=1}^{k'} (1 - x^{-n} y^{-m})} \nonumber \\
    &= Z_\infty \sum_{k,k' = 0}^\infty x^{k N} y^{k' N} \hat{Z}_{(k,k')}.
\end{align}
The terms $x^{k N} Z_\infty \hat{Z}_{(k,0)}$ and $y^{k' N} Z_\infty \hat{Z}_{(0,k')}$ have state-counting interpretations and their charge degeneracies are subject to the lower bound (refined with fugacities) at each $N$. The other terms $\hat{Z}_{(k,k')}$ with $k, k' \neq 0$ do not have state-counting interpretations, because $\hat{Z}_{(k,k')}$ with $k, k' \neq 0$ does not have a power series expression. Computing the right side of \eqref{eq:quarterBPS gge}, e.g. in the limit $x=y=t$, requires one to use analytic properties of $\hat{Z}_{(k,k')}$ in fugacities $x, y$ to cancel poles that are present in that limit. The spectrum on the RHS at each total value of $k+k'$ therefore does not possess a state-counting interpretation. Wall-crossing phenomena due to the analytic properties of such expansions were discussed in the context of superconformal indices in \cite{Lee:2022vig}. We discuss the resolution in this example in Section \ref{sec:basicexamples}.

\section{Basic examples} \label{sec:basicexamples}

We provide some simple examples of Koszul-Tate resolutions of finite $N$ modules, such as those for a $U(N)$-gauged free fermion, $\frac{1}{4}$-BPS sector of free $\Nfour$ SYM, and the $\frac{1}{4}$-BPS chiral ring of $\Nfour$ SYM. Higher ghosts appear in the free $\frac{1}{4}$-BPS sector and the $\frac{1}{4}$-BPS chiral ring.

We hope it is clear to the reader that generalization to more interesting examples, such as $\frac{1}{16}$-BPS sector of $\mathcal{N}=4$ SYM or thermal partition function of pure Yang-Mills, is certainly possible in principle. However, this would require a significant improvement in the algorithm and would benefit from a better understanding of their string duals for comparison of the results.

\subsection{Enumerating relations} \label{subsec:enumeraterelations}

We can enumerate the trace relations in a $U(N)$ gauge theory by defining the function \cite{Concini:2017}
\begin{equation}
\Delta (M_1,\cdots,M_{N+1}) = \sum_{\sigma \in S_{N+1}} \chi_{\underbrace{\scriptstyle (1,1,\cdots,1)}_{N+1}} (\sigma) \, (M_1)_{i_1}^{i_{\sigma(1)}} (M_2)_{i_2}^{i_{\sigma(2)}} \cdots (M_{N+1})_{i_{N+1}}^{i_{\sigma(N+1)}}.
\end{equation}
which consists of $N+1$ matrices $M_1, M_2, \cdots, M_{N+1}$. Here, $\chi_R(\sigma)$ is the character of the symmetric group $S_{N+1}$ in the antisymmetric representation $R=(1,1,\cdots,1)$. As a consequence of the Cayley-Hamilton theorem, the following relation
\begin{equation}
\Delta (M_1,\cdots,M_{N+1}) = 0.
\end{equation}
holds identically for any set of $N \times N$ matrices $M_i$, where each $M_i$ may be grassmann even or odd. Trace relations in a $U(N)$ gauge theory can be found by considering replacements of $M_i$ with (products of) matrix-valued fields in the gauge theory.\footnote{See for example \cite{Dempsey:2022uie} for a recent study using the Cayley-Hamilton theorem to determine trace relations.} Since there are infinitely many trace relations, the set of trace relations must be computed in practice with a cutoff in the charges.

There is a large redundancy in the set of trace relations obtained from making substitutions in $\Delta (M_1,\cdots,M_{N+1}) = 0$. We find the minimal set of trace relation generators and the minimal free resolution of $\cH_N$, up to a truncation in the charge values, by using a Gr\"{o}bner basis algorithm in a computer program \cite{M2}.

\subsection{Examples} \label{subsec:examples}

\paragraph{$U(N)$-gauged fermion}

Consider a free fermion $\psi$ in the adjoint representation of the gauge group $U(N)$. To simplify matters, we consider the Hilbert space built from gauge-invariant operators without derivatives, i.e. traces of an $N \times N$ fermion matrix. Writing down trace relations is straightforward, and the Koszul-Tate resolution of this example can be worked out analytically. Taking the energy of the fermion to be $1/2$, the thermal partition function of the system with anti-periodic boundary conditions is
\begin{equation}
\Tr_{\cH_N} e^{-\beta H} = \prod_{n=1}^N (1 + q^{n-\frac{1}{2}}),
\end{equation}
and with periodic boundary conditions is
\begin{equation}
\Tr_{\cH_N} (-1)^F e^{-\beta H} = \prod_{n=1}^N (1 - q^{n-\frac{1}{2}}),
\end{equation}
where $q = e^{-\beta}$. The fermion number grading $(-1)^F$ here should be distinguished from the heavy ghost grading $(-1)^k$ in the resolution. It will be interesting to see the effect of having both $(-1)^F$ and $(-1)^k$ for the periodic case.

Let us find the trace relations and the resolution. From the cyclicity of the trace, one finds that traces containing even powers of $\psi$ vanish identically, independently of $N$:
\begin{equation}
\Tr \psi^{2n} = 0 \quad \mathrm{for} \ n\geq 1.
\end{equation}
So the large $N$ ring $R$ is
\begin{equation}
R = \mathbb{C}[\Tr \psi, \Tr \psi^3, \Tr \psi^5, \cdots].
\end{equation}
The trace relations at $N$ are
\begin{equation}
\Tr \psi^{2N + 1} = \Tr \psi^{2N + 3} = \Tr \psi^{2N + 5} = \cdots = 0.
\end{equation}
We can find the Koszul-Tate resolution of the finite $N$ module
\begin{equation}
\cH_N = \cH_\infty / \langle \Tr \psi^{2N + 1}, \Tr \psi^{2N + 3}, \Tr \psi^{2N + 5}, \cdots \rangle_R
\end{equation}
by introducing ghosts $\varphi$ and a differential $\hcQ = [\,\cdot\,,\hQ]$ on operators satisfying
\begin{equation}
\hcQ (\varphi_{-2N- a} ) = \Tr \psi^{2N + a}
\end{equation}
where $a = 1,3,5, \cdots$. The anti-commuting differential $\hQ$ in a graded free resolution must commute with all quantum numbers other than the ghost degree of the resolution, so the ghosts $\varphi_{-2N -a}$ are commuting operators with $F = 1$ mod $\mathbb{Z}_2$ and $H = N + \frac{a}{2}$.\footnote{$\varphi$ may both commute and have $F = 1$ mod $\mathbb{Z}_2$ because $F$ is defined through other quantum numbers such as Lorentz spin in higher dimensions.} The differential is nilpotent $\hcQ^2 = 0$ and acts via the graded Leibniz rule on multi-ghost operators:
\begin{align}
    \hcQ ( \varphi_{-2N-a_1} \cdots \varphi_{-2N-a_k}) &= \sum_{i=1}^k ( \varphi_{-2N-a_1} \cdots \hcQ(\varphi_{-2N-a_1}) \cdots \varphi_{-2N-a_k} ) \nonumber \\
    &= \sum_{i=1}^k \Tr \psi^{2N + a_i} ( \varphi_{-2N-a_1} \cdots \varphi_{-2N-a_{i-1}} \varphi_{-2N-a_{i+1}} \cdots \varphi_{-2N-a_k} ).
\end{align}

There are no higher ghosts in this example. This statement can be checked by computing the Lefschetz trace formula using only the multi-ghost states (with coefficients in $R$)
\begin{equation}
\varphi_{-2N-a_1} \varphi_{-2N-a_2} \cdots \varphi_{-2N-a_k} |0 \rangle \in \cV_k
\end{equation}
where $a = 1,3,5, \cdots$ with $1 \leq a_1 \leq a_2 \leq \cdots \leq a_k < \infty$ and no higher ghosts. Let us compute the ghost partition functions by taking traces with respect to the free $R$-modules $\cV_k$. The $k$-ghost partition function without $(-1)^F$ is
\begin{align}
    \Tr_{\cV_k} e^{-\beta H} &= \prod_{n=1}^\infty (1 + q^{n-\frac{1}{2}}) \sum_{\substack{1 \leq a_1 \leq \cdots \leq a_k < \infty \\ a_i \ \mathrm{odd}}} q^{N + \frac{a_1}{2}} q^{N + \frac{a_2}{2}} \cdots q^{N + \frac{a_k}{2}} \nonumber \\
    &= \prod_{n=1}^\infty (1 + q^{n-\frac{1}{2}}) \frac{q^{k N} q^{\frac{k}{2}}}{\prod_{m=1}^k(1-q^m)},
\end{align}
and with $(-1)^F$ is
\begin{align}
    \Tr_{\cV_k} (-1)^F e^{-\beta H} &= \prod_{n=1}^\infty (1 - q^{n-\frac{1}{2}}) \sum_{\substack{1 \leq a_1 \leq \cdots \leq a_k < \infty \\ a_i \ \mathrm{odd}}} (-1)^k q^{N + \frac{a_1}{2}} q^{N + \frac{a_2}{2}} \cdots q^{N + \frac{a_k}{2}} \nonumber \\
    &= \prod_{n=1}^\infty (1 - q^{n-\frac{1}{2}}) \frac{(-1)^k q^{k N} q^{\frac{k}{2}}}{\prod_{m=1}^k(1-q^m)}.
\end{align}
In the latter, $(-1)^F$ evaluates to $(-1)^k$ because each $\varphi$ has $F = 1$ mod $\mathbb{Z}_2$ and heavy ghost number $1$.

We find that the Lefschetz trace formula is satisfied in both cases. For the fermion $\psi$ with anti-periodic boundary conditions one obtains
\begin{equation}
\prod_{n=1}^N (1 + q^{n-\frac{1}{2}}) = \prod_{n=1}^\infty (1 + q^{n-\frac{1}{2}}) \sum_{k=0}^\infty (-1)^k q^{k N} \frac{q^{\frac{k}{2}}}{\prod_{m=1}^k(1-q^m)}
\end{equation}
and with periodic boundary conditions one obtains
\begin{equation}
\prod_{n=1}^N (1 - q^{n-\frac{1}{2}}) = \prod_{n=1}^\infty (1 - q^{n-\frac{1}{2}}) \sum_{k=0}^\infty q^{k N} \frac{q^{\frac{k}{2}}}{\prod_{m=1}^k(1-q^m)}.
\end{equation}
It is interesting that $(-1)^k$ cancels with $(-1)^F$ in the latter. If a string dual of a $U(N)$-gauged fermion can be formulated, this suggests that D-branes in the anti-periodic case behave with fermionic statistics while D-branes in the periodic case behave as bosons.

\paragraph{Free $\frac{1}{4}$-BPS sector}

Let us consider the $\frac{1}{4}$-BPS sector of free $U(N)$ $\Nfour$ super Yang-Mills, which consists of gauge-invariants constructed from two $N \times N$ scalars $X$ and $Y$ \cite{Dolan:2007rq}. Starting with examples involving two matrices, the trace relations become rather involved. We find the trace relations at low values of $N$ up to a truncation in the total dimension of the adjoint fields $X$ and $Y$. Then we reduce the set of trace relations to a minimal set of generators using \cite{M2}.

The partition function of this system is
\begin{equation}
Z_{N}(x,y) = \frac{1}{N!} \oint \prod_{a=1}^N \frac{d\sigma_a}{2 \pi i \sigma_a} \frac{\prod_{a>b} (1 - \sigma_a \sigma_b^{-1})(1 - \sigma_b \sigma_a^{-1})}{\prod_{a,b=1}^N (1 - x \sigma_a \sigma_b^{-1}) (1 - y \sigma_a \sigma_b^{-1})}.
\end{equation}
where $x$ and $y$ are fugacities for charges associated to fields $X$ and $Y$. The large $N$ ring
\begin{equation}
R = \mathbb{C}[\Tr X, \Tr Y, \Tr X^2, \Tr X Y, \cdots ]
\end{equation}
is the ring of polynomials in all single traces constructed from $X,Y$, where $X,Y$ do not commute. The large $N$ partition function is
\begin{equation}
Z_\infty(x,y) = \prod_{n=1}^\infty \frac{1}{1 - x^n - y^n}.
\end{equation}

Let us compute $Z_N$ at low values of $N$ to set some expectations regarding the higher relations. We can read off the single-trace generators of the free $\frac{1}{4}$-BPS sector at low values of $N$ from $Z_N$. The partition function for $N=1,2$ are
\begin{align}
    Z_1 &= \frac{1}{(1-x)(1-y)} \nonumber \\
    Z_2 &= \frac{1}{(1-x)(1-y)(1-x^2)(1-x y)(1-y^2)}.
\end{align}
This suggests that $\cH_{N=1}$ is a Fock space built from the generators $\Tr X, \, \Tr Y$, and that $\cH_{N=2}$ is a Fock space built from
\begin{equation}
\Tr X,\, \Tr Y,\, \Tr X^2,\, \Tr X Y,\, \Tr Y^2.
\end{equation}
However, the partition function at $N=3$
\begin{equation}
Z_3 = \frac{(1-x^6 y^6)}{\left(\prod_{n=1}^3(1-x^n)(1-y^n)\right)(1-x y)(1-x^2 y)(1-x y^2)(1-x^2 y^2)(1-x^3 y^3)}
\end{equation}
says that $\cH_{N=3}$ cannot be a Fock space built from single trace generators \cite{Hanany:2007zz}, because there exists a relation of degree $(6,6)$ in $X,Y$ fields among the generators
\begin{equation}
\Tr X,\, \Tr Y,\, \Tr X^2,\, \Tr X Y,\, \Tr Y^2, \Tr X^3,\, \Tr X^2 Y,\, \Tr X Y^2,\, \Tr Y^3,\, \Tr X^2 Y^2,\, \Tr X^3 Y^3.
\end{equation}
One can think of other traces, e.g. $\Tr X Y X Y$, $\Tr X Y X^2 Y^2$, and $\Tr X Y X Y X Y$, at degrees $(2,2)$ and $(3,3)$ as having been eliminated by expressing them in terms of $\Tr X^2 Y^2$ and $\Tr X^3 Y^3$ using the trace relations at $N=3$. The analysis in this paragraph amounts to using the trace relations that were present at each $N$ to reduce the ring generators to a finite set, but we would like to work over the large $N$ ring $R$ and find whether there exist non-trivial relations between trace relations.

Let $\cV_k^N$ be free $R$-modules consisting a minimal Koszul-Tate resolution $\hcK$ of $\cH_N$. We define the following, which is the Hilbert series of $\cV_k^N$:
\begin{equation}
Z_{\cV_k^N} = \Tr_{\cV_k^N} \left[ x^{\cC_X} y^{\cC_Y} \right].
\end{equation}
By enumerating trace relations, we find that the generators of $\cV_1^N$ that map to trace relations possess the following total charges:
\begin{align}
    \frac{Z_{\cV_1^{N=1}}}{Z_\infty} &= 3 t^2 + 4 t^3 + 6 t^4 + 8 t^5 + 14 t^6 + 20 t^7 + 36 t^8 + 60 t^9 + 108 t^{10} + 188 t^{11} + 352 t^{12} + \cdots \nonumber \\
    \frac{Z_{\cV_1^{N=2}}}{Z_\infty} &= 4 t^3 + 6 t^4 + 8 t^5 + 14 t^6 + 20 t^7 + 36 t^8 + 60 t^9 + 108 t^{10} + 188 t^{11} + 352 t^{12} + \cdots \nonumber \\
    \frac{Z_{\cV_1^{N=3}}}{Z_\infty} &= 5 t^4 + 8 t^5 + 13 t^6 + 20 t^7 + 36 t^8 + 60 t^9 + 108 t^{10} + 188 t^{11} + 352 t^{12} + \cdots
\end{align}
where we specialized the fugacities to $x = y = t$.

We would like to find whether there exist higher relations between trace relations. We can do so by checking whether the Lefschetz trace formula is satisfied by multi-ghost states constructed only from single-ghost generators with the above charges. We use Polya theory \cite{Sundborg:1999ue,Polyakov:2001af,Aharony:2003sx} to multi-particle the anti-commuting ghosts of $\cV_1^N$:
\begin{equation}
\exp\left[ - \sum_{n=1}^\infty \frac{\mu^n}{n} \frac{Z_{\cV_1^{N}}}{Z_\infty}(t^n) \right] = \sum_{k=0}^\infty (-1)^k \mu^k M_{k, N} (t)
\end{equation}
where the series $M_{k, N} (t)$ contains charges of states with total heavy ghost number $k$. If the $M_{k, N} (t)$ above satisfy
\begin{equation} \label{eq:polya question}
\frac{Z_N(t)}{Z_\infty(t)} \ \stackrel{?}{=} \ \sum_{k=0}^\infty (-1)^k M_{k, N} (t),
\end{equation}
it means that there are no ghosts-for-ghosts.

Up to total charge $12$, we find that \eqref{eq:polya question} is satisfied for $N=1,2$. There is a discrepancy at $N=3$ of
\begin{equation} \label{eq:free quarterBPS question}
\frac{Z_3(t)}{Z_\infty(t)} - \sum_{k=0}^\infty (-1)^k M_{k, 3} (t) = - t^{12} + O(t^{13})
\end{equation}
which indicates that the Koszul-Tate resolution at $N=3$ involves a higher ghost state at total charge $12$. The minus sign, and the fact that the $M_{k=3, 3} (t)$ is the final term to contribute to \eqref{eq:free quarterBPS question} up to charge $12$, suggests the higher ghost $\chi^{(3)}|0\rangle \in \cV_{k=3}^{N=3}$ of charge $12$ maps under $\hQ$ to a linear combination of $\chi_{A_1}^{(1)} \chi_{A_2}^{(1)} |0\rangle$ with coefficients in $R$.

\paragraph{$\frac{1}{4}$-BPS chiral ring}

A similar computation may be done in the $\frac{1}{4}$-BPS chiral ring of $U(N)$ $\Nfour$ SYM, which involves adjoint fields $X$ and $Y$ subject to a superpotential constraint $[X,Y]=0$. The large $N$ ring $R$ consists of traces of commuting fields:
\begin{equation}
R = \mathbb{C}[ \Tr X^n Y^m ]
\end{equation}
where $n,m=0,1,2, \cdots$ and $(n,m) \neq (0,0)$. The grand canonical partition function of the $\frac{1}{4}$-BPS chiral ring is
\begin{equation}
\cZ(p; x,y) = \sum_{N=0}^\infty p^N Z_N(x,y) = \prod_{n,m=0}^\infty \frac{1}{1 - p \, x^n y^m}
\end{equation}
and the large $N$ partition function is
\begin{equation}
Z_\infty (x,y) = \prod_{\substack{n,m=0 \\ (n,m) \neq (0,0)}}^\infty \frac{1}{1 - x^n y^m}.
\end{equation}

One finds non-trivial results also starting at $N=3$, where the generators of $\cV_1^{N=3}$ that map to trace relations possess the following total charges:
\begin{equation}
\frac{Z_{\cV_1^{N=3}}}{Z_\infty} = 5 t^4 + 8 t^5 + 10 t^6 + 8 t^7 + 9 t^8 + 10 t^9 + 11 t^{10} + \cdots
\end{equation}
There is a discrepancy in \eqref{eq:polya question} at $N=3$:
\begin{equation}
\frac{Z_3(t)}{Z_\infty(t)} - \sum_{k=0}^\infty (-1)^k M_{k, 3} (t) = 3 t^{8} + O(t^{9}),
\end{equation}
which indicates that three ghost-for-ghost states $\chi^{(2)}|0\rangle \in \cV_{k=2}^{N=3}$ exist in the minimal Koszul-Tate resolution at charge $8$. It would be nice to understand the higher ghosts in this context by comparing the large $N$ spectrum of $\cG_k$ in Section \ref{subsubsec:largeN} at fixed $\cC_Y$ charges with the spectrum of normalizable states constructed from higher $S^3$ modes on D3 giants.

\section{Discussion} \label{sec:discussion}

We conclude with a partial list of open questions:
\begin{enumerate}
    \item A path integral realization of the Koszul-Tate resolution implementing trace relations would be conceptually enlightening. This may help understand how structures discussed in this work are realized in the bulk string field theory in the $\alpha' \to \infty$ limit.
    \item To make the bulk grading and the Hilbert space structure manifest, we used canonical quantization to study the states of bulk D-branes. It would be nice to understand how the grading $(-1)^k$ arises in the gravitational path integral in the presence of Euclidean D-branes.
    \item Trace relations exist at finite $\lambda$. How does the structure of the Koszul-Tate resolution change when one turns on the coupling, and what additional ingredients are required? How do we describe the dynamics of the heavy ghost states, and how are properties of black holes encoded therein? How does the bulk radial coordinate emerge at finite $\lambda$? It would also be nice to understand the recent progress in $\frac{1}{16}$-BPS states at weak coupling \cite{Chang:2022mjp,Choi:2022caq,Choi:2023znd,Budzik:2023vtr,Chang:2023zqk} using tools presented in this work.
    \item More intricate finite $N$ relations appear in gauge theories with (anti-)fundamental and adjoint fields or in gauge theories with surface defects with (anti-)fundamental degrees of freedom. It would be nice to gain an algebraic understanding of various branes in anti-de Sitter space in terms of relations at finite $N$.
    \item It would be important to have a sharper bulk understanding of the higher ghosts (e.g. as strings on multiply-wrapped branes, etc.). Along these lines, it would be helpful to generalize the computation of normalizable states in Section \ref{subsec:localization} to that with higher $S^3$ modes and fluctuations in $\AdS$ directions.
    \item It would be nice to understand the relationship between our work and works on the K-theory classification of D-branes \cite{Green:1996dd,Minasian:1997mm,Witten:1998cd,Freed:2000tt,Maldacena:2001xj,Witten:2000cn}, higher spin holography \cite{Klebanov:2002ja,Sezgin:2002rt,Giombi:2009wh,Gaberdiel:2010pz}, tensionless strings \cite{Gaberdiel:2018rqv,Eberhardt:2018ouy,Eberhardt:2019ywk,Gaberdiel:2021jrv}, and works relating holography and Koszul duality \cite{Costello:2016mgj,Costello:2017fbo,Ishtiaque:2018str,Costello:2018zrm,Costello:2020jbh}.
    \item  The holographic interpretation of the differential $\hQ$ is an instanton which maps a state on $k$ wrapped D-branes to a linear combination of states on $k-1$ wrapped D-branes with coefficients in the traces (i.e. closed strings). What are the gauge theory predictions for the tunneling amplitudes between vacua labelled by different D-branes, as well as the effect of turning on $\lambda$ on this amplitude?
\end{enumerate}

\subsection*{Acknowledgments}

I thank D. Belayneh, M. R. Gaberdiel, D. Gaiotto, M. Heydeman, L. V. Iliesiu, S. Komatsu, W. Li, E. J. Martinec, H. Murali, S. S. Pufu, S. Raghavendran, P. Vieira, and E. Witten for helpful discussions. I am especially grateful to D. Gaiotto for many valuable suggestions and careful reading of the draft. I thank the organizers of the Precision Holography workshop at CERN and the organizers of the QFT \& Strings seminar at ETH Z\"{u}rich for the opportunity to present early versions of this work. I am thankful for the hospitality of Institut f\"{u}r Theoretische Physik at ETH Z\"{u}rich, where parts of this work was done.

I am supported by the Perimeter Institute for Theoretical Physics and in part by the NSERC Discovery Grant program and the Simons Collaboration on Confinement and QCD Strings. Research at Perimeter Institute is supported in part by the Government of Canada through the Department of Innovation, Science and Economic Development Canada and by the Province of Ontario through the Ministry of Colleges and Universities.

\bibliographystyle{./JHEP}
\bibliography{references.bib}

\end{document}